\documentclass[12pt,letterpaper,oneside]{article}
\usepackage[margin=1in]{geometry}
\usepackage{natbib}
\usepackage{float}
\usepackage{graphicx}
\usepackage{amsmath, amssymb, hyperref, ulem}
\usepackage{xcolor, authblk}
\usepackage{multirow}
\usepackage{mwe}
\usepackage{subfig}

\usepackage{lineno}
\usepackage{caption}
\title{Phase Neural Operator for Multi-Station Picking of Seismic Arrivals}

\author[1]{Hongyu Sun \thanks{Corresponding author: hongyu-sun@outlook.com}}
\author[1]{Zachary E. Ross}
\author[1, 2]{Weiqiang Zhu}
\author[3]{Kamyar Azizzadenesheli}

\affil[1]{Seismological Laboratory, California Institute of Technology, 1200 E. California Blvd., Pasadena, CA 91125}

\affil[2]{Current address: Berkeley Seismological Laboratory, University of California, Berkeley, 307 McCone Hall, Berkeley, CA 94720}

\affil[3]{Nvidia Corporation, 2788 San Tomas Expressway,
Santa Clara, CA 95051}

\date{}

\begin{document}

\maketitle

\textbf{Key Points:}
\begin{itemize}
  \item We introduce a multi-station phase picking algorithm, PhaseNO, that is based on a new machine learning paradigm called Neural Operator.
  \item PhaseNO can use data from any number of stations arranged in any arbitrary geometry to pick phases across the input stations simultaneously.
  \item By leveraging the spatial and temporal contextual information, PhaseNO achieves superior performance over leading baseline algorithms.
\end{itemize}

\section*{Abstract}

Seismic wave arrival time measurements form the basis for numerous downstream applications. State-of-the-art approaches for phase picking use deep neural networks to annotate seismograms at each station independently, yet human experts annotate seismic data by examining the whole network jointly. Here, we introduce a general-purpose network-wide phase picking algorithm based on a recently developed machine learning paradigm called Neural Operator. Our model, called PhaseNO, leverages the spatio-temporal contextual information to pick phases simultaneously for any seismic network geometry. This results in superior performance over leading baseline algorithms by detecting many more earthquakes, picking more phase arrivals, while also greatly improving measurement accuracy. Following similar trends being seen across the domains of artificial intelligence, our approach provides but a glimpse of the potential gains from fully-utilizing the massive seismic datasets being collected worldwide.

\section*{Plain Language Summary}
Earthquake monitoring often involves measuring arrival times of P- and S-waves of earthquakes from continuous seismic data. With the advancement of artificial intelligence, state-of-the-art phase picking methods use deep neural networks to examine seismic data from each station independently; this is in stark contrast to the way that human experts annotate seismic data, in which waveforms from the whole network containing multiple stations are examined simultaneously. With the performance gains of single-station algorithms approaching saturation, it is clear that meaningful future advances will require algorithms that can naturally examine data for entire networks at once. Here we introduce a multi-station phase picking algorithm based on a recently developed machine learning paradigm called Neural Operator. Our algorithm, called PhaseNO, leverages the spatial-temporal information of earthquake signals from an input seismic network with arbitrary geometry. This results in superior performance over leading baseline algorithms by detecting many more earthquakes, picking many more seismic wave arrivals, yet also greatly improving measurement accuracy.

\section{Introduction}

Seismic phase detection and picking are fundamental tasks in earthquake seismology, where the aim is to identify earthquakes in the continuous data and measure the arrival times of seismic waves. Historically, human seismic analysts manually labeled earthquake signals and the arrival times of seismic phases by looking for coherent wavefronts on multiple stations and then picking the onset times of P and S waves at each station. Such analysis, however, is subjective, time-consuming, and prone to errors. Considerable effort has been dedicated to developing accurate, automatic, and timely earthquake detection methods, such as short-term average/long-term average \citep{withers_comparison_1998}, template matching \citep{gibbons_detection_2006,shelly_non-volcanic_2007}, and fingerprint and similarity threshold \citep{yoon_earthquake_2015}. Recent advances in deep learning have greatly improved the accuracy and efficiency of automatic phase picking algorithms \citep{perol_convolutional_2018,ross_generalized_2018,zhu_phasenet_2018,mousavi_cred_2019,zhu_deep_2019,zhou_hybrid_2019,wang_deep_2019,dokht_seismic_2019,mousavi_earthquake_2020,yeck_leveraging_2021,xiao_siamese_2021,zhu_endend_2022,feng_edgephase_2022,munchmeyer_which_2022,Johnson_2022}. However, the single-station detection strategy used in most of the machine-learning detection algorithms can result in failure to detect events with weak amplitude, or mistakenly detect local noise signals with emergence pulses. Indeed, the performance gains of single-station neural phase pickers have rapidly saturated, leading to the question of where the next breakthroughs in phase picking will come from.

Across the various domains of artificial intelligence, such as natural language processing and computer vision, the largest gains in performance have come from (i) using ever-larger datasets with increasingly detailed labeling/prediction tasks, (ii) making sense of unlabeled data, and (iii) incorporating powerful model architectures (e.g. transformers) that are capable of learning to extract information from these very complex datasets. Translating these successes to the phase picking problem would similarly require formulating the problem more generally, in which the goal is to output phase picks only after examining the seismic data for all available sensors in a network. To accomplish such a general formulation, new models are needed that can naturally consider the spatial and temporal context on a variable arrangement of sensors. Although strategies have been proposed to handle the irregular seismic network geometry for earthquake source characterization \cite{van2020automated,zhang2022spatiotemporal},  earthquake early warning \cite{munchmeyer2021transformer,bloemheuvel2022graph}, and seismic phase association \cite{mcbrearty2023earthquake}, a generalized network-based phase picker remains an open question.

In this paper, we introduce such an approach for general purpose network-wide earthquake detection and phase picking. Our algorithm, called Phase Neural Operator (PhaseNO), builds on Neural Operators \citep{kovachki_neural_2023}, a recent advance of deep learning models that operate directly on functions rather than finite dimensional vectors. PhaseNO learns infinite dimensional function representations of seismic wavefields across the network, allowing us to accurately measure the arrival times of different phases jointly at multiple stations with arbitrary geometry. We evaluate our approach on real-world seismic datasets and compare its performance with state-of-the-art phase picking methods. We demonstrate that PhaseNO outperforms leading baseline algorithms by detecting many more earthquakes, picking many more phase arrivals, yet also greatly improving measurement accuracy. Overall, our approach demonstrates the power of leveraging both temporal and spatial information for seismic phase picking and improving earthquake monitoring systems. 

\section{Method: Phase Neural Operator}

We introduce an operator learning model for network-wide phase picking (see Text S1). PhaseNO is designed to learn an operator between infinite-dimensional function spaces on a bounded physical domain. The input function is a seismic wavefield observed at some arbitrary collection of points in space and time, $f(x,y,t)$, and the output function is a probability mask $g(x,y,t)$ that indicates the likelihood of P- and S-wave arrivals at each point $(x,y,t)$. A powerful advantage of Neural Operators over classical Neural Networks is that they are discretization-invariant, meaning that the input and output functions can be discretized on a different (arbitrary) mesh every time a solution is to be evaluated, without having to re-train the model. This critical property allows for Neural Operators to be evaluated at any point within the input physical domain, enabling phase picking on a dynamic seismic network with different geometries.

We combine two types of Neural Operators to naturally handle the mathematical structure of seismic network data. For the temporal information, we use Fourier Neural Operator (FNO) layers \citep{li_fourier_2020}, which are ideal for cases in which the domain is sure to be discretized on a regular mesh, because fast Fourier transforms are used to quickly compute a solution. Since seismograms are mostly sampled regularly in time, FNO can efficiently process and encode seismograms. For the spatial information, our sensors are generally not on a regular mesh, and so we instead use Graph Neural Operators \citep[GNO,][]{li_neural_2020} to model the relationship of seismic waveforms at different stations. This type of neural operator is naturally able to work with irregular sensors, as it uses message passing \citep{gilmer_neural_2017} to aggregate features from multiple stations and construct an operator with kernel integration.

\begin{figure}
\centering
\includegraphics[width=\textwidth]{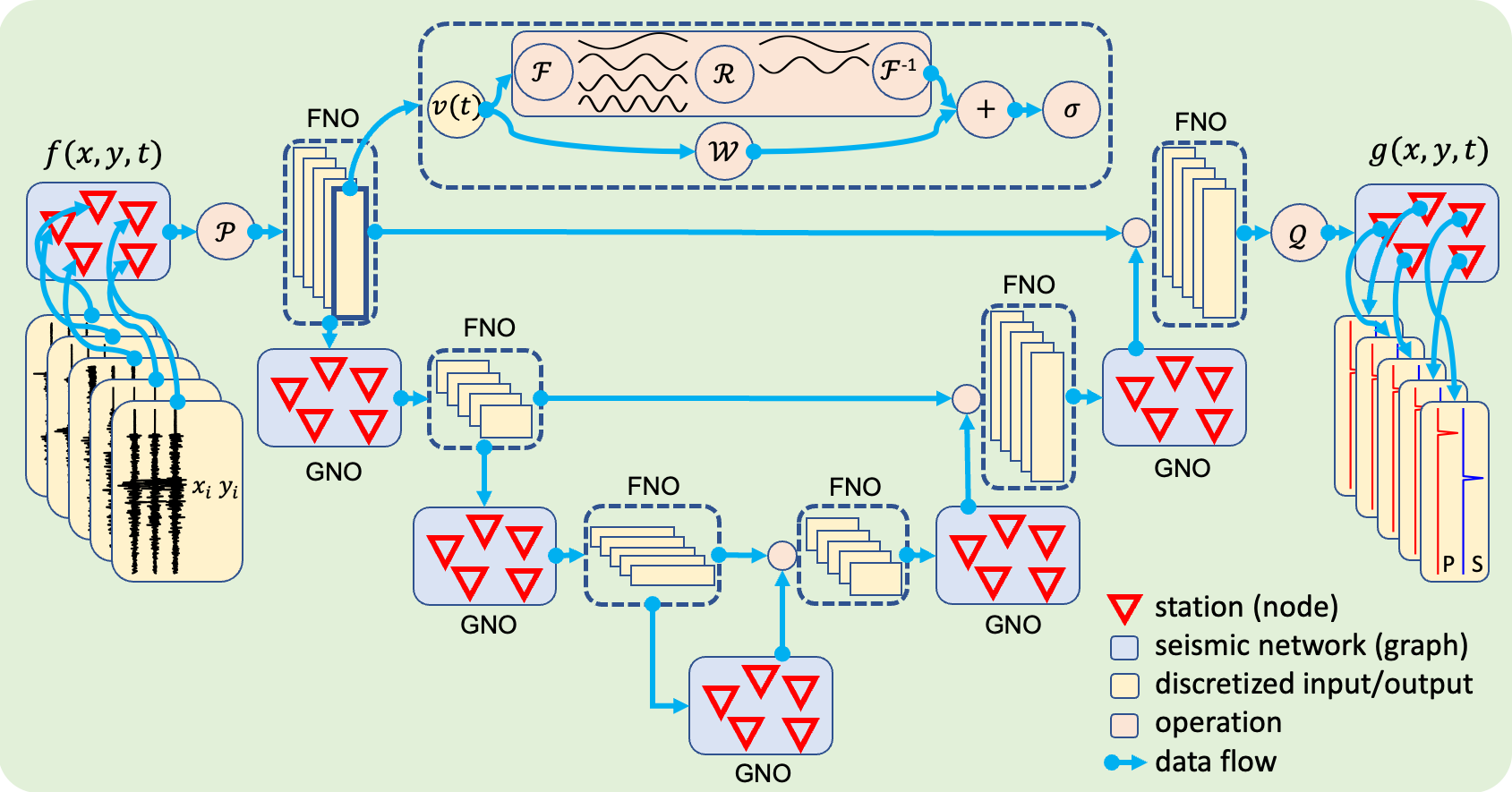}
\caption{\textbf{PhaseNO architecture.} The model consists of multiple FNO and GNO layers that are sequentially connected and repeated. $\mathcal{P}$ and $\mathcal{Q}$ are up- and down-projections parameterized by neural networks. The model uses seismograms from a seismic network containing multiple stations with an arbitrary geometry as the input and predicts the probabilities of P-phase and S-phase arrival times for all input stations. Station locations are encoded as two channels of the input, in addition to three channels carrying the three-component waveforms. The relative locations $(x_i, y_i)$ between stations can be used to learn weights as edge features in a graph (see Text S2).}
\label{fig:model}
\end{figure}

Figure~\ref{fig:model} summarizes the PhaseNO architecture. The model is composed of multiple blocks of operator layers in which FNO and GNO are sequentially connected and repeated several times, allowing for sufficient communications and exchange of spatiotemporal information between all stations in a seismic network. Skip connections are used to connect the blocks, resulting in a U-shape architecture. The skip connection directly concatenates FNO results on the left part of the model with GNO results on the right without going through deep layers, which improves convergence and allows for deeper, more overparameterized models.

\begin{figure}
\centering
\includegraphics[width=\textwidth]{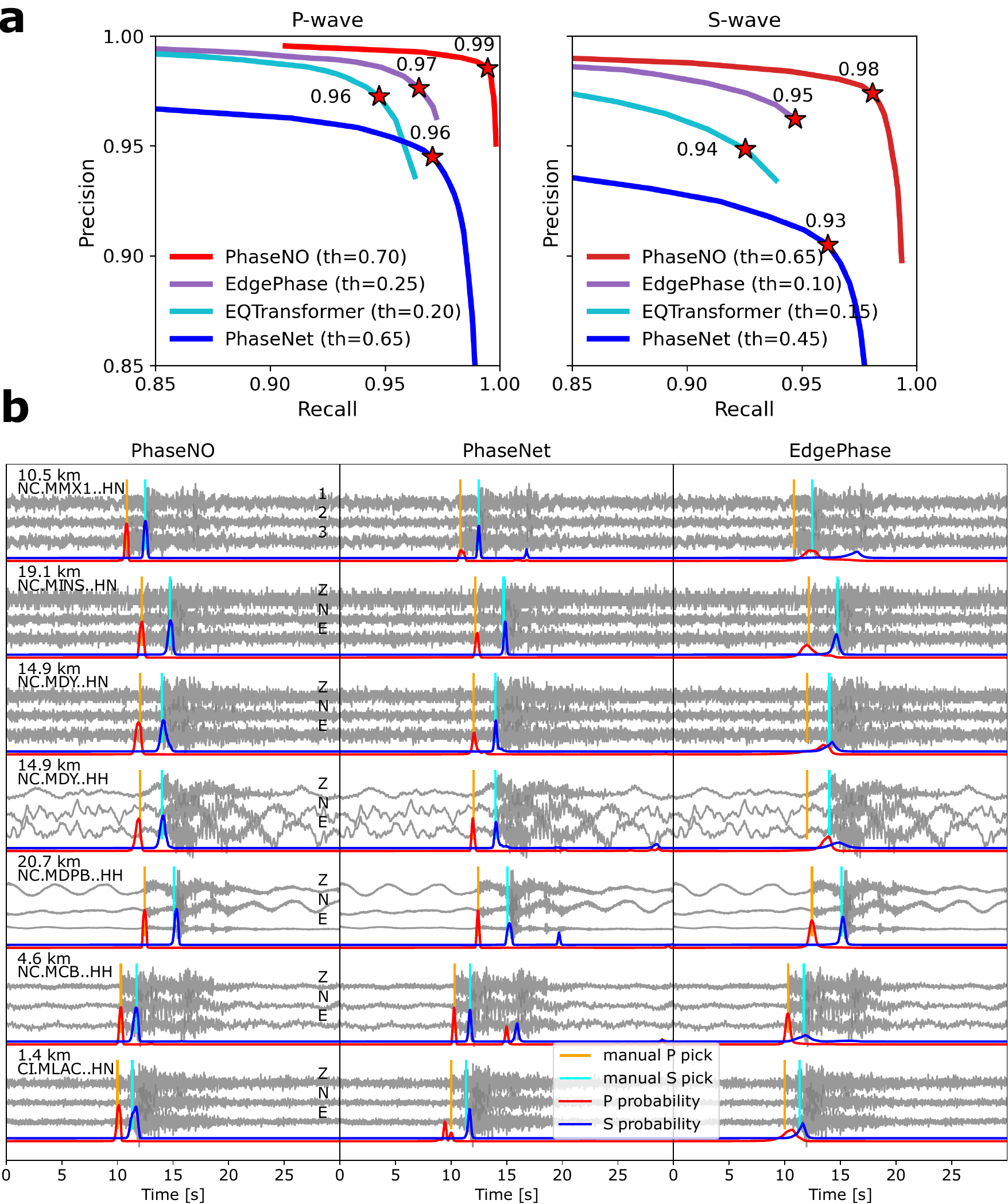}
\caption{\textbf{Performance evaluation on the NCEDC2020 test dataset.} (a) Precision-recall curves. The best threshold (th) for each model on this test dataset is selected based on the maximum F1 scores (stars labeled on the curves) that models achieve (Figure \ref{figs1:threshold}). (b) Event nc71112909 with a magnitude of 0.43. The station name and epicentral distance are shown on waveforms.} 
\label{fig:precision_recall_test_waveforms}
\end{figure}

\section{Results}

\subsection{Performance evaluation}

In this study, we benchmark the performance of PhaseNO against three leading baseline models (see Text S3): EQTransformer \citep{mousavi_earthquake_2020}, PhaseNet \citep{zhu_phasenet_2018}, and EdgePhase \citep{feng_edgephase_2022}. We trained PhaseNO on an earthquake dataset from the Northern California Earthquake Data Center (NCEDC) spanning the period 1984-2019 (see Text S4), i.e., the same training dataset as PhaseNet. We evaluated PhaseNO and each baseline model on an out of sample test dataset for the period 2020 containing 43,700 P/S picks of 5,769 events. We choose the time window for each sample based on their pre-trained models: 30 s for PhaseNO and PhaseNet, and 60 s for EQTransformer and EdgePhase. Positions of picks are randomly placed in the middle 30 s of the time window. For all of the models, P- and S-picks were determined from peaks in the predicted probability distributions by setting a pre-determined threshold. Each model used a distinct threshold as the one maximizing the F1 score to ensure the models compared under their best conditions (Figure~\ref{fig:precision_recall_test_waveforms}a; Figure~\ref{figs1:threshold}).

Our method results in the highest F1 scores for both P- and S-waves, being 0.99 and 0.98 respectively. This is in addition to having the highest optimal thresholds (0.70 for P and 0.65 for S) of all the models tested (Table~\ref{tab:scores}). Given that similar labeling strategies were used for training the baselines (Gaussian for PhaseNet and triangular for the other models), a higher threshold indicates that PhaseNO has a higher confidence level for detecting and picking seismic arrivals than other methods. When true picks are unavailable to determine the optimal threshold for a particular test dataset based on F1 scores (i.e., the trade-off between correct and false phases), PhaseNO is able to minimize false detection and give more picks compared with other methods with the same pre-determined threshold. The two single station picking models, PhaseNet and EQTransformer, have similar F1 scores, but the former has higher recall and the latter has higher precision. EdgePhase is built on EQTransformer and has better performance in terms of the precision-recall curves. However, the phase picks are less precise in terms of time residuals (Table~\ref{tab:scores}; Figure~\ref{figs2:hist}). PhaseNO detects more true positives, fewer false negatives, and fewer false positive picks than the other deep-learning models at almost all signal-to-noise levels (Figure~\ref{fig:snr_histogram}). Despite generating more picks, PhaseNO results in the smallest mean absolute error for both P and S phases. Overall, PhaseNO achieves the best performance on all six metrics, with one minor exception. The standard deviation of P phase residuals for PhaseNO is 0.01 s (one time step) larger than PhaseNet. It should be noted that the newly detected phases by PhaseNO are likely to be more challenging cases as their signal-to-noise levels are lower, and thus result in slightly increased standard deviation.

We compare the predicted probability distributions of each neural phase picker for several representative events (Figures~\ref{fig:precision_recall_test_waveforms}b, \ref{figs4:test_waveforms}, and \ref{figs5:test_waveforms}). PhaseNO works very well on different event magnitudes, instrument types, and waveform shapes. PhaseNet generates some false positive picks that are removed by multi-station methods (PhaseNO and EdgePhase); however, EdgePhase also generates many false negatives. Through exchanging temporal and spatial information multiple times, PhaseNO effectively prevents false picks while improving the detection ability of true picks. PhaseNO successfully finds picks on low SNR waveforms by leveraging contextual information from other stations.

S-phases generally exist in the coda of P-phases and are more challenging to find. Thus, more labeling errors from human analysts are expected on S phases than P phases. For instance, in Figure \ref{figs4:test_waveforms}a, three of the models generate consistent S picks, but the predicted peaks are systematically offset from the manual picks on this event. For these example cases, PhaseNO shows significant improvement in S-phase picking and generates higher probabilities than the other methods. Moreover, the width of the picks predicted by PhaseNO may represent the degree of difficulty in picking the phases from the waveforms, even though the same label width is used in baselines and our method. Picks with high probabilities may have wider distribution if the waveforms have low SNR. Also, our model can handle the waveforms with more than one pick existing in a sample (Figure \ref{figs5:test_waveforms}).

\subsection{Application to the 2019 Ridgecrest earthquake sequence}

We tested the detection performance and generalization ability of PhaseNO on the 2019 Ridgecrest earthquake sequence. We downloaded continuous waveform data for EH, HH, and HN sensors for the period 4 July, 2019 (15:00:00) to 10 July, 2019 (00:00:00) at 20 SCSN stations, which is a total of 36 distinct sensors. Each of these sensors is treated as a distinct node in the graph, even if they are co-located (Table \ref{tab:stations}). Waveform data are divided into hourly streams with a sampling rate of 100 Hz. This is a challenging dataset due to the overlap of numerous events. Since no ground-truth catalog is available for the continuous data, we evaluated our results by comparing them with catalogs produced by SCSN, PhaseNet, and two template matching studies \citep{ross_hierarchical_2019,shelly_high-resolution_2020}.

\begin{figure}
\centering
\includegraphics[width=\textwidth]{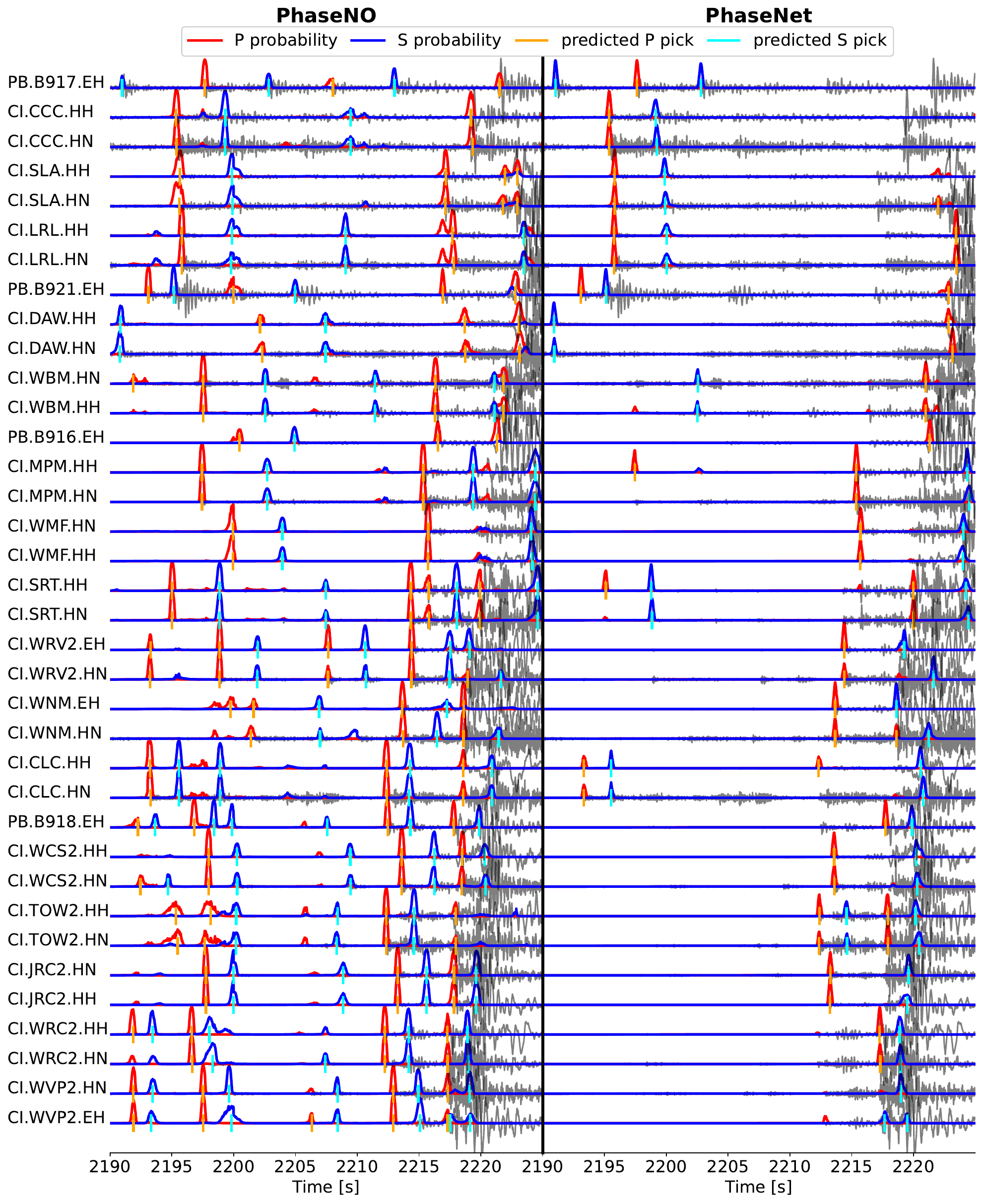}
\caption{Example results for a 35-s window during the 2019 Ridgecrest earthquake sequence.} 
\label{fig:ridgecrest_waveform_070604_2190}
\end{figure}

\begin{figure}
\centering
\includegraphics[width=\textwidth]{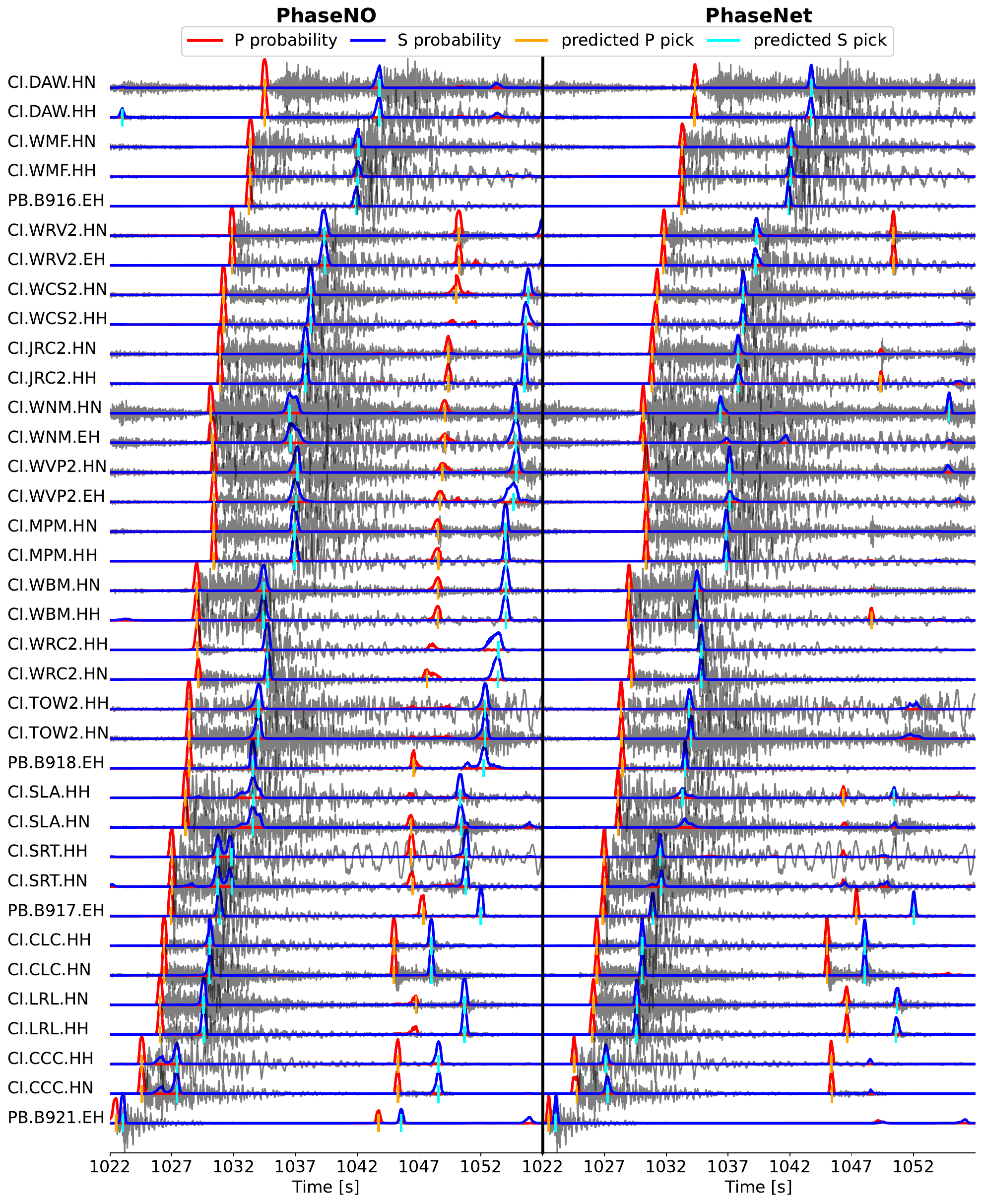}
\caption{Example results for a short window during the 2019 Ridgecrest earthquake sequence. More examples can be found in the Supporting Information (Figures \ref{fig:ridgecrest_waveform_070606_240} – \ref{figs9:ridgecrest_waveform_070702_3420})}
\label{fig:ridgecrest_waveform_maintext2}
\end{figure}

We first divided the entire seismic network into two parts and constructed two graphs for every hour of data, due to the increased computational cost with the number of nodes in a graph (Figure~\ref{figs5:cost}). The 36 nodes were randomly divided into two graphs with 18 nodes. Continuous data were cut into a 30-s time window with an overlap of 10 s, resulting in 180 predictions for one-hour data on 18 nodes. After preprocessing, PhaseNO predicted the probabilities of earthquake phases on 18 nodes at once. We compare representative waveforms with probabilities predicted by PhaseNO and PhaseNet (Figures \ref{fig:ridgecrest_waveform_070604_2190}, \ref{fig:ridgecrest_waveform_maintext2}, and \ref{fig:ridgecrest_waveform_070606_240} – \ref{figs9:ridgecrest_waveform_070702_3420}). Both models show great generalization ability, as these waveforms were recorded outside of the training region. Our model works very well on continuous data, especially when there is more than one event in a 30-s time window, when the event is located at any position of the window, and when the waveform has different shapes with low SNR. Owing to the learned waveform consistency among multiple stations, PhaseNO detects much more picks with meaningful moveout patterns than PhaseNet. 

After prediction, we determined phase picks using a threshold of 0.3 for both P and S phases. PhaseNO detected 693,266 P and 686,629 S arrival times, while PhaseNet found 542,793 P and 572,991 S arrival times with the same threshold and the same stations. We evaluated the accuracy of the detected picks by comparing the arrival times with manually reviewed picks from SCSN (Figure \ref{figs10:ridgecrest_picks}). The standard deviation of the pick residuals between SCSN and PhaseNO was 0.10 s for P phases and 0.14 s for S phases, calculated from 118,746 P picks and 96,247 S picks. The standard deviation, however, was slightly higher than those with PhaseNet (0.08 s for P from 106,061 picks and 0.13 s for S from 88,438 picks). Since the newly detected picks are more challenging cases with low fidelity, it is reasonable for PhaseNO to show a larger travel time difference.

\begin{figure}
\centering
\includegraphics[width=\textwidth]{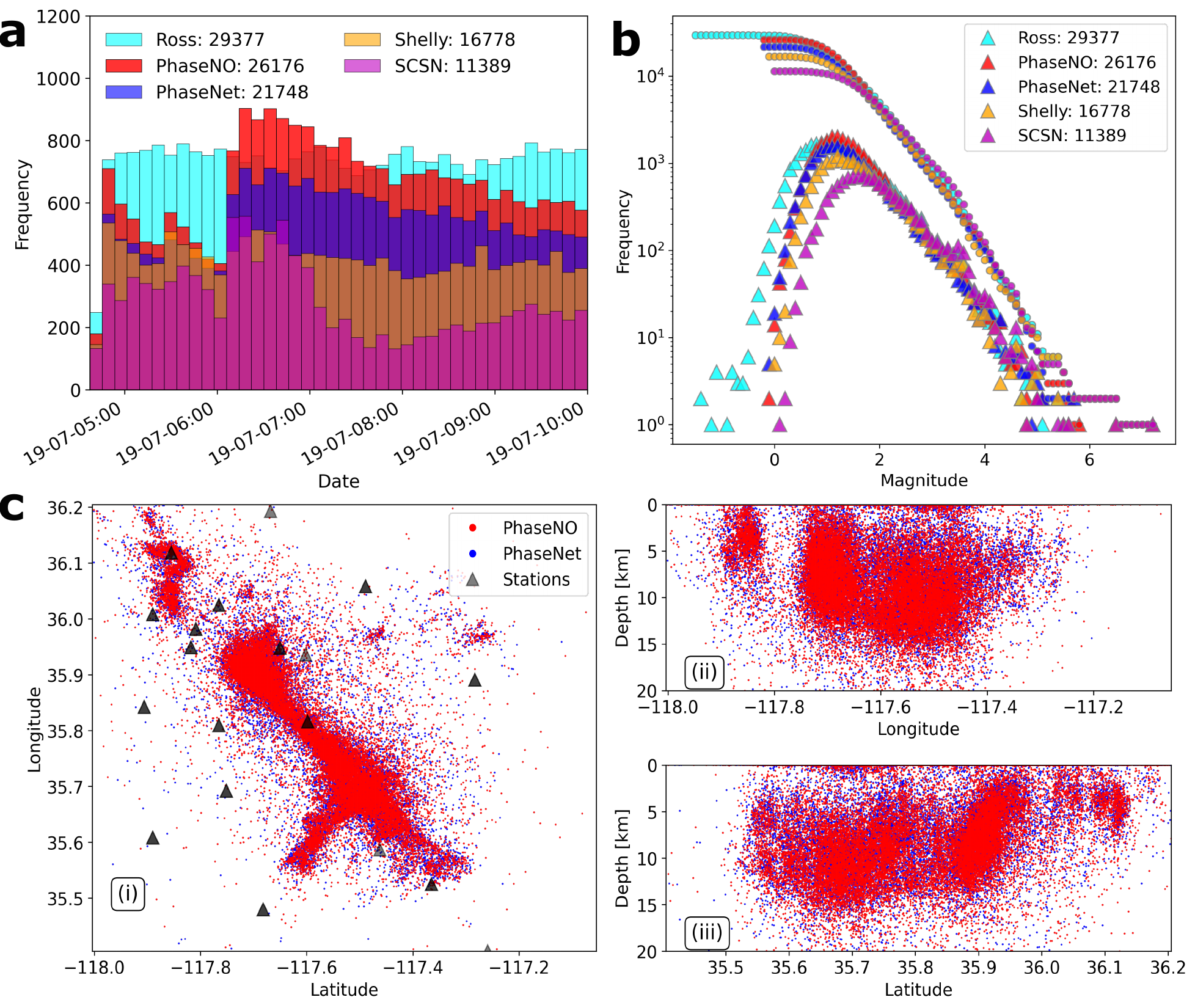}
\caption{\textbf{Comparison of earthquake catalogs of the 2019 Ridgecrest earthquake sequence.} (a) Earthquake number. (b) Frequency-magnitude distributions. (c) Earthquake hypocenters.} 
\label{fig:ridgecrest_catalogs}
\end{figure}

We convert candidate phase detections into events by phase association with GaMMA \citep{zhu_earthquake_2022}. We set a minimum of 17 picks per event to filter out low-quality associations. This results in PhaseNet detecting 21,748 events with 37.54 picks per event, whereas PhaseNO detects 26,176 events with 39.37 picks per event (Figure \ref{fig:ridgecrest_catalogs}a). Many of the unassociated picks are probably a consequence of our strict filtering criteria during association, rather than false detections. With the same association hyperparameters, the additional 4428 events highlight the advancement of PhaseNO for earthquake detection. Despite the increased number of events, PhaseNO shows high detection quality with around two more picks per event compared to PhaseNet, even though they are smaller events in general (Figure \ref{figs11:ridgecrest_events}). GaMMA calculates magnitudes for events detected by PhaseNO and PhaseNet, and they both show linear Gutenberg-Richter distributions (Figure \ref{fig:ridgecrest_catalogs}b). Indeed, our results have fewer microearthquakes than the template matching catalog by \cite{ross_hierarchical_2019}. Since microearthquakes usually have limited propagation ranges and can only be recorded by several stations, they would have been filtered out during association and thus not shown on the frequency-magnitude distribution. Moreover, event locations determined by GaMMA are generally consistent between PhaseNO and PhaseNet catalogs (Figure \ref{fig:ridgecrest_catalogs}c), confirming that the additional events by PhaseNO are reasonable detections of real earthquakes.

Furthermore, we treat the manually reviewed SCSN catalog as a baseline and evaluate how many earthquakes were successfully recovered. We consider that two events are matched if they occur within 3 s from each other. With such criteria, Shelly, Ross et al., and PhaseNet matched around 81$\%$, 86$\%$, and 88$\%$ events, respectively. In comparison, with our strict filtering criteria during association, PhaseNO catalog totaling 26,176 events matched approximately 94$\%$ events in the SCSN catalog (10,673 of 11,389) with additional events, indicating the highest recall score of PhaseNO. PhaseNO consistently detects more events than PhaseNet, SCSN, and Shelly’s template matching catalog \citep{shelly_high-resolution_2020} over time and approaches the number of earthquakes reported by another more detailed template matching catalog \citep{ross_hierarchical_2019}. Moreover, PhaseNO achieves a much more stable detection with the greatest number of events found when the $M_w$ 7.1 mainshock occurred (Figure \ref{fig:ridgecrest_catalogs}a) and with the gradually reduced seismicity rate afterwards, indicating the power of the method to illuminate complex earthquake sequences. Examples of events and associated picks detected by PhaseNO can be found in the Supporting Information (Figures \ref{figs12:phaseno_catalog_event_11262} – \ref{figs14:phaseno_catalog_event_16248}).

It should be noted that our catalog differs from those of the SCSN and template matching catalogs in the number of stations and association algorithms. However, picks from PhaseNO and PhaseNet are detected on the exact same stations and then associated with GaMMA, providing the fairest comparison. Two post-processing hyperparameters, the threshold in phase picking and the minimum number of picks associated with an event, control the total number of earthquakes in a catalog. A lower threshold and a smaller association minimum provide more events, despite likely more false positive events (Table \ref{tab:association}). PhaseNO consistently detects more events than PhaseNet using the same hyperparameters, pointing out the importance of leveraging the spatial information in addition to the temporary information for phase picking.

\section{Discussion and Conclusions}

With a fixed model architecture, PhaseNO can handle seismic networks with arbitrary geometries; we demonstrated this by training on the Northern California Seismic Network and evaluated the model on the Southern California Seismic Network, without retraining. This is a critical property of the Neural Operator class of models, which can learn in infinite dimensions.

PhaseNO shows several distinctive characteristics in terms of network design. Compared to most of the currently popular detection algorithms (deep-learning or traditional methods), PhaseNO mimics human learning and decision making by using context from the whole seismic network, rather than seismograms at a single station. By consulting information and searching for consistent waveforms from surrounding stations, PhaseNO greatly improves phase picking on low SNR data, especially S phases that usually are hidden in the coda of P phases. 

Apart from the characteristics in the spatial domain, PhaseNO has a unique ability to identify phases from temporal information. The well-known transformer architecture that has brought about major successes in natural language processing \citep{vaswani_attention_2017} can be viewed as a special case of Neural Operators \citep{kovachki_neural_2023}. Just as EQTransformer uses an attention mechanism to investigate global dependencies, PhaseNO supervises the global features with kernel integrals in space and time. Like PhaseNet, PhaseNO adopts a U-shape architecture with skip connections, which improves model convergence and allows for a deeper model design with greater expressiveness.

Compared to EdgePhase, a multi-station picking model, our model uses multiple GNO layers, a type of Neural Operator that allows for kernel integration over the network to extract rich spatial features. Each GNO layer is inserted between two FNO layers, forcing the exchange of information between spatial and temporal domains. We also encode station locations as node features to weight the message constructed between nodes. Additionally, instead of building a graph based on geographic distances and only selecting neighboring nodes within a certain distance from the target node, we construct a graph using all nodes in a seismic network. All these modifications contribute to maximizing the usage of spatial features for phase picking. 

A major limitation to PhaseNO, however, is the dependence of memory usage on the number of stations in one prediction. Spatial information is exchanged between all pairs of nodes in a graph; therefore, the computational cost scales quadratically with the number of nodes, with complexity $O(n^2)$. Hence, we suggest selecting a subset of stations from the entire large seismic network for one prediction until all stations have been processed before moving to the next time segment of continuous data, like the procedure described in the Ridgecrest example. If the seismic network covers a wide range of areas, we may select stations based on k-means clustering \citep{lloyd1982least}. In this way, we can greatly accelerate the prediction procedure and save memory usage, particularly when there are many stations and when the computational resources are limited. 

\section{Open Research}

Version v1.0.0 of PhaseNO and the pre-trained model are preserved at \cite{sun_2023_10224301}. The training and test data are from Northern California Earthquake Data Center (NCEDC), doi:10.7932/NCEDC. The data of the 2019 Ridgecrest earthquake sequence can be accessed from Southern California Earthquake Data Center, doi:10.7909/C3WD3xH1, and Plate Boundary Observatory Borehole Seismic Network, doi:10.7932/NCEDC.

\section{Acknowledgements}
We thank the Editor Daoyuan Sun, Christopher W Johnson and an anonymous reviewer for constructive comments on the manuscript. ZER is grateful to the David and Lucile Packard Foundation for supporting this study through a Packard Fellowship.

\bibliographystyle{plainnat}
\bibliography{phaseno.bib}

\begin{thebibliography}{39}
\providecommand{\natexlab}[1]{#1}
\providecommand{\url}[1]{\texttt{#1}}
\expandafter\ifx\csname urlstyle\endcsname\relax
  \providecommand{\doi}[1]{doi: #1}\else
  \providecommand{\doi}{doi: \begingroup \urlstyle{rm}\Url}\fi

\bibitem[Bloemheuvel et~al.(2022)Bloemheuvel, van~den Hoogen, Jozinovi{\'c}, Michelini, and Atzmueller]{bloemheuvel2022graph}
Stefan Bloemheuvel, Jurgen van~den Hoogen, Dario Jozinovi{\'c}, Alberto Michelini, and Martin Atzmueller.
\newblock Graph neural networks for multivariate time series regression with application to seismic data.
\newblock \emph{International Journal of Data Science and Analytics}, pages 1--16, 2022.

\bibitem[Dokht et~al.(2019)Dokht, Kao, Visser, and Smith]{dokht_seismic_2019}
Ramin M.~H. Dokht, Honn Kao, Ryan Visser, and Brindley Smith.
\newblock Seismic {Event} and {Phase} {Detection} {Using} {Time}–{Frequency} {Representation} and {Convolutional} {Neural} {Networks}.
\newblock \emph{Seismological Research Letters}, 90\penalty0 (2A):\penalty0 481--490, March 2019.
\newblock ISSN 0895-0695, 1938-2057.
\newblock \doi{10.1785/0220180308}.
\newblock URL \url{https://pubs.geoscienceworld.org/ssa/srl/article/90/2A/481/568235/Seismic-Event-and-Phase-Detection-Using}.

\bibitem[Feng et~al.(2022)Feng, Mohanna, and Meng]{feng_edgephase_2022}
Tian Feng, Saeed Mohanna, and Lingsen Meng.
\newblock {EdgePhase}: {A} {Deep} {Learning} {Model} for {Multi}‐{Station} {Seismic} {Phase} {Picking}.
\newblock \emph{Geochemistry, Geophysics, Geosystems}, 23\penalty0 (11), November 2022.
\newblock ISSN 1525-2027, 1525-2027.
\newblock \doi{10.1029/2022GC010453}.
\newblock URL \url{https://onlinelibrary.wiley.com/doi/10.1029/2022GC010453}.

\bibitem[Gibbons and Ringdal(2006)]{gibbons_detection_2006}
Steven~J. Gibbons and Frode Ringdal.
\newblock The detection of low magnitude seismic events using array-based waveform correlation.
\newblock \emph{Geophysical Journal International}, 165\penalty0 (1):\penalty0 149--166, April 2006.
\newblock ISSN 0956540X, 1365246X.
\newblock \doi{10.1111/j.1365-246X.2006.02865.x}.
\newblock URL \url{https://academic.oup.com/gji/article-lookup/doi/10.1111/j.1365-246X.2006.02865.x}.

\bibitem[Gilmer et~al.(2017)Gilmer, Schoenholz, Riley, Vinyals, and Dahl]{gilmer_neural_2017}
Justin Gilmer, Samuel~S. Schoenholz, Patrick~F. Riley, Oriol Vinyals, and George~E. Dahl.
\newblock Neural {Message} {Passing} for {Quantum} {Chemistry}, June 2017.
\newblock URL \url{http://arxiv.org/abs/1704.01212}.

\bibitem[Hendrycks and Gimpel(2020)]{hendrycks_gaussian_2020}
Dan Hendrycks and Kevin Gimpel.
\newblock Gaussian {Error} {Linear} {Units} ({GELUs}), July 2020.
\newblock URL \url{http://arxiv.org/abs/1606.08415}.

\bibitem[Johnson and Johnson(2022)]{Johnson_2022}
Christopher~W Johnson and Paul~A. Johnson.
\newblock {EQDetect}: Earthquake phase arrivals and first motion polarity applying deep learning.
\newblock apr 2022.
\newblock \doi{10.1002/essoar.10511191.1}.
\newblock URL \url{https://doi.org/10.1002%2Fessoar.10511191.1}.

\bibitem[Kovachki et~al.(2023)Kovachki, Li, Liu, Azizzadenesheli, Bhattacharya, Stuart, and Anandkumar]{kovachki_neural_2023}
Nikola Kovachki, Zongyi Li, Burigede Liu, Kamyar Azizzadenesheli, Kaushik Bhattacharya, Andrew Stuart, and Anima Anandkumar.
\newblock Neural {Operator}: {Learning} {Maps} {Between} {Function} {Spaces} {With} {Applications} to {PDEs}.
\newblock \emph{Journal of Machine Learning Research}, 24\penalty0 (89):\penalty0 1--97, 2023.
\newblock URL \url{http://jmlr.org/papers/v24/21-1524.html}.

\bibitem[Li et~al.(2020{\natexlab{a}})Li, Kovachki, Azizzadenesheli, Liu, Bhattacharya, Stuart, and Anandkumar]{li_fourier_2020}
Zongyi Li, Nikola Kovachki, Kamyar Azizzadenesheli, Burigede Liu, Kaushik Bhattacharya, Andrew Stuart, and Anima Anandkumar.
\newblock Fourier {Neural} {Operator} for {Parametric} {Partial} {Differential} {Equations}.
\newblock 2020{\natexlab{a}}.
\newblock \doi{10.48550/ARXIV.2010.08895}.
\newblock URL \url{https://arxiv.org/abs/2010.08895}.

\bibitem[Li et~al.(2020{\natexlab{b}})Li, Kovachki, Azizzadenesheli, Liu, Bhattacharya, Stuart, and Anandkumar]{li_neural_2020}
Zongyi Li, Nikola Kovachki, Kamyar Azizzadenesheli, Burigede Liu, Kaushik Bhattacharya, Andrew Stuart, and Anima Anandkumar.
\newblock Neural {Operator}: {Graph} {Kernel} {Network} for {Partial} {Differential} {Equations}, March 2020{\natexlab{b}}.
\newblock URL \url{http://arxiv.org/abs/2003.03485}.

\bibitem[Lloyd(1982)]{lloyd1982least}
Stuart Lloyd.
\newblock Least squares quantization in pcm.
\newblock \emph{IEEE transactions on information theory}, 28\penalty0 (2):\penalty0 129--137, 1982.

\bibitem[McBrearty and Beroza(2023)]{mcbrearty2023earthquake}
Ian~W McBrearty and Gregory~C Beroza.
\newblock Earthquake phase association with graph neural networks.
\newblock \emph{Bulletin of the Seismological Society of America}, 113\penalty0 (2):\penalty0 524--547, 2023.

\bibitem[Mousavi et~al.(2019{\natexlab{a}})Mousavi, Sheng, Zhu, and Beroza]{mousavi_stanford_2019}
S.~Mostafa Mousavi, Yixiao Sheng, Weiqiang Zhu, and Gregory~C. Beroza.
\newblock {STanford} {EArthquake} {Dataset} ({STEAD}): {A} {Global} {Data} {Set} of {Seismic} {Signals} for {AI}.
\newblock \emph{IEEE Access}, 7:\penalty0 179464--179476, 2019{\natexlab{a}}.
\newblock ISSN 2169-3536.
\newblock \doi{10.1109/ACCESS.2019.2947848}.
\newblock URL \url{https://ieeexplore.ieee.org/document/8871127/}.

\bibitem[Mousavi et~al.(2019{\natexlab{b}})Mousavi, Zhu, Sheng, and Beroza]{mousavi_cred_2019}
S.~Mostafa Mousavi, Weiqiang Zhu, Yixiao Sheng, and Gregory~C. Beroza.
\newblock {CRED}: {A} {Deep} {Residual} {Network} of {Convolutional} and {Recurrent} {Units} for {Earthquake} {Signal} {Detection}.
\newblock \emph{Scientific Reports}, 9\penalty0 (1):\penalty0 10267, July 2019{\natexlab{b}}.
\newblock ISSN 2045-2322.
\newblock \doi{10.1038/s41598-019-45748-1}.
\newblock URL \url{https://www.nature.com/articles/s41598-019-45748-1}.

\bibitem[Mousavi et~al.(2020)Mousavi, Ellsworth, Zhu, Chuang, and Beroza]{mousavi_earthquake_2020}
S.~Mostafa Mousavi, William~L. Ellsworth, Weiqiang Zhu, Lindsay~Y. Chuang, and Gregory~C. Beroza.
\newblock Earthquake transformer—an attentive deep-learning model for simultaneous earthquake detection and phase picking.
\newblock \emph{Nature Communications}, 11\penalty0 (1):\penalty0 3952, August 2020.
\newblock ISSN 2041-1723.
\newblock \doi{10.1038/s41467-020-17591-w}.
\newblock URL \url{https://www.nature.com/articles/s41467-020-17591-w}.

\bibitem[M{\"u}nchmeyer et~al.(2021)M{\"u}nchmeyer, Bindi, Leser, and Tilmann]{munchmeyer2021transformer}
Jannes M{\"u}nchmeyer, Dino Bindi, Ulf Leser, and Frederik Tilmann.
\newblock The transformer earthquake alerting model: A new versatile approach to earthquake early warning.
\newblock \emph{Geophysical Journal International}, 225\penalty0 (1):\penalty0 646--656, 2021.

\bibitem[Münchmeyer et~al.(2022)Münchmeyer, Woollam, Rietbrock, Tilmann, Lange, Bornstein, Diehl, Giunchi, Haslinger, Jozinović, Michelini, Saul, and Soto]{munchmeyer_which_2022}
Jannes Münchmeyer, Jack Woollam, Andreas Rietbrock, Frederik Tilmann, Dietrich Lange, Thomas Bornstein, Tobias Diehl, Carlo Giunchi, Florian Haslinger, Dario Jozinović, Alberto Michelini, Joachim Saul, and Hugo Soto.
\newblock Which {Picker} {Fits} {My} {Data}? {A} {Quantitative} {Evaluation} of {Deep} {Learning} {Based} {Seismic} {Pickers}.
\newblock \emph{Journal of Geophysical Research: Solid Earth}, 127\penalty0 (1), January 2022.
\newblock ISSN 2169-9313, 2169-9356.
\newblock \doi{10.1029/2021JB023499}.
\newblock URL \url{https://onlinelibrary.wiley.com/doi/10.1029/2021JB023499}.

\bibitem[Perol et~al.(2018)Perol, Gharbi, and Denolle]{perol_convolutional_2018}
Thibaut Perol, Michaël Gharbi, and Marine Denolle.
\newblock Convolutional neural network for earthquake detection and location.
\newblock \emph{Science Advances}, 4\penalty0 (2):\penalty0 e1700578, February 2018.
\newblock ISSN 2375-2548.
\newblock \doi{10.1126/sciadv.1700578}.
\newblock URL \url{https://www.science.org/doi/10.1126/sciadv.1700578}.

\bibitem[Rahman et~al.(2022)Rahman, Ross, and Azizzadenesheli]{rahman2022u}
Md~Ashiqur Rahman, Zachary~E Ross, and Kamyar Azizzadenesheli.
\newblock {U}-no: {U}-shaped neural operators.
\newblock \emph{arXiv preprint arXiv:2204.11127}, 2022.

\bibitem[Ross et~al.(2018)Ross, Meier, Hauksson, and Heaton]{ross_generalized_2018}
Zachary~E. Ross, Men‐Andrin Meier, Egill Hauksson, and Thomas~H. Heaton.
\newblock Generalized {Seismic} {Phase} {Detection} with {Deep} {Learning}.
\newblock \emph{Bulletin of the Seismological Society of America}, 108\penalty0 (5A):\penalty0 2894--2901, October 2018.
\newblock ISSN 0037-1106, 1943-3573.
\newblock \doi{10.1785/0120180080}.
\newblock URL \url{https://pubs.geoscienceworld.org/ssa/bssa/article/108/5A/2894/546740/Generalized-Seismic-Phase-Detection-with-Deep}.

\bibitem[Ross et~al.(2019)Ross, Idini, Jia, Stephenson, Zhong, Wang, Zhan, Simons, Fielding, Yun, Hauksson, Moore, Liu, and Jung]{ross_hierarchical_2019}
Zachary~E. Ross, Benjamín Idini, Zhe Jia, Oliver~L. Stephenson, Minyan Zhong, Xin Wang, Zhongwen Zhan, Mark Simons, Eric~J. Fielding, Sang-Ho Yun, Egill Hauksson, Angelyn~W. Moore, Zhen Liu, and Jungkyo Jung.
\newblock Hierarchical interlocked orthogonal faulting in the 2019 {Ridgecrest} earthquake sequence.
\newblock \emph{Science}, 366\penalty0 (6463):\penalty0 346--351, October 2019.
\newblock ISSN 0036-8075, 1095-9203.
\newblock \doi{10.1126/science.aaz0109}.
\newblock URL \url{https://www.science.org/doi/10.1126/science.aaz0109}.

\bibitem[Shelly(2020)]{shelly_high-resolution_2020}
David~R. Shelly.
\newblock A {High}-{Resolution} {Seismic} {Catalog} for the {Initial} 2019 {Ridgecrest} {Earthquake} {Sequence}: {Foreshocks}, {Aftershocks}, and {Faulting} {Complexity}.
\newblock \emph{Seismological Research Letters}, 91\penalty0 (4):\penalty0 1971--1978, July 2020.
\newblock ISSN 0895-0695, 1938-2057.
\newblock \doi{10.1785/0220190309}.
\newblock URL \url{https://pubs.geoscienceworld.org/ssa/srl/article/91/4/1971/580481/A-HighResolution-Seismic-Catalog-for-the-Initial}.

\bibitem[Shelly et~al.(2007)Shelly, Beroza, and Ide]{shelly_non-volcanic_2007}
David~R. Shelly, Gregory~C. Beroza, and Satoshi Ide.
\newblock Non-volcanic tremor and low-frequency earthquake swarms.
\newblock \emph{Nature}, 446\penalty0 (7133):\penalty0 305--307, March 2007.
\newblock ISSN 0028-0836, 1476-4687.
\newblock \doi{10.1038/nature05666}.
\newblock URL \url{http://www.nature.com/articles/nature05666}.

\bibitem[Sun(2023)]{sun_2023_10224301}
Hongyu Sun.
\newblock Phase{N}{O} (v1.0.0) {Z}enodo [{S}oftware], November 2023.
\newblock URL \url{https://doi.org/10.5281/zenodo.10224301}.

\bibitem[van~den Ende and Ampuero(2020)]{van2020automated}
Martijn~PA van~den Ende and J-P Ampuero.
\newblock Automated seismic source characterization using deep graph neural networks.
\newblock \emph{Geophysical Research Letters}, 47\penalty0 (17):\penalty0 e2020GL088690, 2020.

\bibitem[Vaswani et~al.(2017)Vaswani, Shazeer, Parmar, Uszkoreit, Jones, Gomez, Kaiser, and Polosukhin]{vaswani_attention_2017}
Ashish Vaswani, Noam Shazeer, Niki Parmar, Jakob Uszkoreit, Llion Jones, Aidan~N. Gomez, Lukasz Kaiser, and Illia Polosukhin.
\newblock Attention {Is} {All} {You} {Need}, December 2017.
\newblock URL \url{http://arxiv.org/abs/1706.03762}.

\bibitem[Wang et~al.(2019{\natexlab{a}})Wang, Xiao, Liu, Zhao, and Yao]{wang_deep_2019}
Jian Wang, Zhuowei Xiao, Chang Liu, Dapeng Zhao, and Zhenxing Yao.
\newblock Deep {Learning} for {Picking} {Seismic} {Arrival} {Times}.
\newblock \emph{Journal of Geophysical Research: Solid Earth}, 124\penalty0 (7):\penalty0 6612--6624, July 2019{\natexlab{a}}.
\newblock ISSN 2169-9313, 2169-9356.
\newblock \doi{10.1029/2019JB017536}.
\newblock URL \url{https://onlinelibrary.wiley.com/doi/10.1029/2019JB017536}.

\bibitem[Wang et~al.(2019{\natexlab{b}})Wang, Sun, Liu, Sarma, Bronstein, and Solomon]{wang_dynamic_2019}
Yue Wang, Yongbin Sun, Ziwei Liu, Sanjay~E. Sarma, Michael~M. Bronstein, and Justin~M. Solomon.
\newblock Dynamic {Graph} {CNN} for {Learning} on {Point} {Clouds}.
\newblock \emph{ACM Transactions on Graphics}, 38\penalty0 (5):\penalty0 1--12, October 2019{\natexlab{b}}.
\newblock ISSN 0730-0301, 1557-7368.
\newblock \doi{10.1145/3326362}.
\newblock URL \url{https://dl.acm.org/doi/10.1145/3326362}.

\bibitem[Withers et~al.(1998)Withers, Aster, Young, Beiriger, Harris, Moore, and Trujillo]{withers_comparison_1998}
Mitchell Withers, Richard Aster, Christopher Young, Judy Beiriger, Mark Harris, Susan Moore, and Julian Trujillo.
\newblock A comparison of select trigger algorithms for automated global seismic phase and event detection.
\newblock \emph{Bulletin of the Seismological Society of America}, 88\penalty0 (1):\penalty0 95--106, February 1998.
\newblock ISSN 1943-3573, 0037-1106.
\newblock \doi{10.1785/BSSA0880010095}.
\newblock URL \url{https://pubs.geoscienceworld.org/bssa/article/88/1/95/102726/A-comparison-of-select-trigger-algorithms-for}.

\bibitem[Xiao et~al.(2021)Xiao, Wang, Liu, Li, Zhao, and Yao]{xiao_siamese_2021}
Zhuowei Xiao, Jian Wang, Chang Liu, Juan Li, Liang Zhao, and Zhenxing Yao.
\newblock Siamese {Earthquake} {Transformer}: {A} {Pair}‐{Input} {Deep}‐{Learning} {Model} for {Earthquake} {Detection} and {Phase} {Picking} on a {Seismic} {Array}.
\newblock \emph{Journal of Geophysical Research: Solid Earth}, 126\penalty0 (5), May 2021.
\newblock ISSN 2169-9313, 2169-9356.
\newblock \doi{10.1029/2020JB021444}.
\newblock URL \url{https://onlinelibrary.wiley.com/doi/10.1029/2020JB021444}.

\bibitem[Yeck et~al.(2021)Yeck, Patton, Ross, Hayes, Guy, Ambruz, Shelly, Benz, and Earle]{yeck_leveraging_2021}
William~Luther Yeck, John~M. Patton, Zachary~E. Ross, Gavin~P. Hayes, Michelle~R. Guy, Nick~B. Ambruz, David~R. Shelly, Harley~M. Benz, and Paul~S. Earle.
\newblock Leveraging {Deep} {Learning} in {Global} 24/7 {Real}-{Time} {Earthquake} {Monitoring} at the {National} {Earthquake} {Information} {Center}.
\newblock \emph{Seismological Research Letters}, 92\penalty0 (1):\penalty0 469--480, January 2021.
\newblock ISSN 0895-0695, 1938-2057.
\newblock \doi{10.1785/0220200178}.
\newblock URL \url{https://pubs.geoscienceworld.org/ssa/srl/article/92/1/469/591060/Leveraging-Deep-Learning-in-Global-24-7-Real-Time}.

\bibitem[Yoon et~al.(2015)Yoon, O’Reilly, Bergen, and Beroza]{yoon_earthquake_2015}
Clara~E. Yoon, Ossian O’Reilly, Karianne~J. Bergen, and Gregory~C. Beroza.
\newblock Earthquake detection through computationally efficient similarity search.
\newblock \emph{Science Advances}, 1\penalty0 (11):\penalty0 e1501057, December 2015.
\newblock ISSN 2375-2548.
\newblock \doi{10.1126/sciadv.1501057}.
\newblock URL \url{https://www.science.org/doi/10.1126/sciadv.1501057}.

\bibitem[Zhang et~al.(2022{\natexlab{a}})Zhang, Liu, Feng, Wang, and Zhu]{zhang2022loc}
Miao Zhang, Min Liu, Tian Feng, Ruijia Wang, and Weiqiang Zhu.
\newblock {LOC}-{FLOW}: {A}n end-to-end machine learning-based high-precision earthquake location workflow.
\newblock \emph{Seismological Society of America}, 93\penalty0 (5):\penalty0 2426--2438, 2022{\natexlab{a}}.

\bibitem[Zhang et~al.(2022{\natexlab{b}})Zhang, Reichard-Flynn, Zhang, Hirn, and Lin]{zhang2022spatiotemporal}
Xitong Zhang, Will Reichard-Flynn, Miao Zhang, Matthew Hirn, and Youzuo Lin.
\newblock Spatiotemporal graph convolutional networks for earthquake source characterization.
\newblock \emph{Journal of Geophysical Research: Solid Earth}, 127\penalty0 (11):\penalty0 e2022JB024401, 2022{\natexlab{b}}.

\bibitem[Zhou et~al.(2019)Zhou, Yue, Kong, and Zhou]{zhou_hybrid_2019}
Yijian Zhou, Han Yue, Qingkai Kong, and Shiyong Zhou.
\newblock Hybrid {Event} {Detection} and {Phase}‐{Picking} {Algorithm} {Using} {Convolutional} and {Recurrent} {Neural} {Networks}.
\newblock \emph{Seismological Research Letters}, 90\penalty0 (3):\penalty0 1079--1087, May 2019.
\newblock ISSN 0895-0695, 1938-2057.
\newblock \doi{10.1785/0220180319}.
\newblock URL \url{https://pubs.geoscienceworld.org/ssa/srl/article/90/3/1079/569837/Hybrid-Event-Detection-and-PhasePicking-Algorithm}.

\bibitem[Zhu et~al.(2019)Zhu, Peng, McClellan, Li, Yao, Li, and Fang]{zhu_deep_2019}
Lijun Zhu, Zhigang Peng, James McClellan, Chenyu Li, Dongdong Yao, Zefeng Li, and Lihua Fang.
\newblock Deep learning for seismic phase detection and picking in the aftershock zone of 2008 {M7}.9 {Wenchuan} {Earthquake}.
\newblock \emph{Physics of the Earth and Planetary Interiors}, 293:\penalty0 106261, August 2019.
\newblock ISSN 00319201.
\newblock \doi{10.1016/j.pepi.2019.05.004}.
\newblock URL \url{https://linkinghub.elsevier.com/retrieve/pii/S0031920118301407}.

\bibitem[Zhu and Beroza(2018)]{zhu_phasenet_2018}
Weiqiang Zhu and Gregory~C Beroza.
\newblock {PhaseNet}: {A} {Deep}-{Neural}-{Network}-{Based} {Seismic} {Arrival} {Time} {Picking} {Method}.
\newblock \emph{Geophysical Journal International}, October 2018.
\newblock ISSN 0956-540X, 1365-246X.
\newblock \doi{10.1093/gji/ggy423}.
\newblock URL \url{https://academic.oup.com/gji/advance-article/doi/10.1093/gji/ggy423/5129142}.

\bibitem[Zhu et~al.(2022{\natexlab{a}})Zhu, McBrearty, Mousavi, Ellsworth, and Beroza]{zhu_earthquake_2022}
Weiqiang Zhu, Ian~W. McBrearty, S.~Mostafa Mousavi, William~L. Ellsworth, and Gregory~C. Beroza.
\newblock Earthquake {Phase} {Association} using a {Bayesian} {Gaussian} {Mixture} {Model}.
\newblock \emph{Journal of Geophysical Research: Solid Earth}, 127\penalty0 (5), May 2022{\natexlab{a}}.
\newblock ISSN 2169-9313, 2169-9356.
\newblock \doi{10.1029/2021JB023249}.
\newblock URL \url{http://arxiv.org/abs/2109.09008}.

\bibitem[Zhu et~al.(2022{\natexlab{b}})Zhu, Tai, Mousavi, Bailis, and Beroza]{zhu_endend_2022}
Weiqiang Zhu, Kai~Sheng Tai, S.~Mostafa Mousavi, Peter Bailis, and Gregory~C. Beroza.
\newblock An {End}‐{To}‐{End} {Earthquake} {Detection} {Method} for {Joint} {Phase} {Picking} and {Association} {Using} {Deep} {Learning}.
\newblock \emph{Journal of Geophysical Research: Solid Earth}, 127\penalty0 (3):\penalty0 e2021JB023283, March 2022{\natexlab{b}}.
\newblock ISSN 2169-9313, 2169-9356.
\newblock \doi{10.1029/2021JB023283}.
\newblock URL \url{https://agupubs.onlinelibrary.wiley.com/doi/10.1029/2021JB023283}.

\end{thebibliography}

\newpage
\setcounter{page}{1}
\linespread{1.0}

\section*{Supporting Information for ``Phase Neural Operator for Multi-Station Picking of Seismic Arrivals"}

Hongyu Sun$^1$, Zachary E. Ross$^1$, Weiqiang Zhu$^{1,2}$, and Kamyar Azizzadenesheli$^3$ \\ 

\noindent 1. \textit{Seismological Laboratory, California Institute of Technology, 1200 E. California Blvd., Pasadena, CA 91125}

\noindent 2. \textit{Current address: Berkeley Seismological Laboratory, University of California, Berkeley, 307 McCone Hall, Berkeley, CA 94720}

\noindent 3. \textit{Nvidia Corporation, 2788 San Tomas Expressway,
Santa Clara, CA 95051}

\vspace{10mm}

\noindent \textbf{\Large Contents}

\begin{description}
   \item[Text S1] Neural Operators.
   \item[Text S2] Encoding station locations as node features.
   \item[Text S3] Baselines for model evaluation.
   \item[Text S4] Training and test datasets.
   \item[Text S5] Training details and evaluation metrics.
   \item[Text S6] More discussions.
   \item[Table S1] Phase picking performance on the test dataset. The best score is highlighted in bold.
   \item[Table S2] List of stations used for PhaseNO and PhaseNet in the 2019 Ridgecrest earthquake sequence.
   \item[Table S3] Comparison of association results for picks detected by PhaseNO and PhaseNet in the 2019 Ridgecrest earthquake sequence. 
   \item[Table S4] Comparison of training datasets.
   \item[Figure S1] Comparison of the threshold sensitivity for PhaseNO, PhaseNet, EQTransformer and EdgePhase on the test dataset.
   \item[Figure S2] Comparison of picking errors of deep-learning models on the test dataset.
   \item[Figure S3] Phase picking performance as a function of noise level.
   \item[Figures S4 and S5] Results of representative events in the test dataset.
   \item[Figure S6] PhaseNO scaling properties as a function of the number of nodes in one graph.
   \item[Figures S7 - S11] Representative waveforms from the 2019 Ridgecrest earthquake sequence.
   \item[Figure S12] Travel time difference between SCSN picks and picks detected by PhaseNO or PhaseNet in the 2019 Ridgecrest earthquake sequence.
   \item[Figure S13] Comparison of the association quality of GaMMA using picks detected by PhaseNO and PhaseNet in the 2019 Ridgecrest earthquake sequence through the number of associated picks in one event.
   \item[Figure S14] An example of an M 0.825 event detected by PhaseNO.
   \item[Figure S15] An example of an M 1.528 event detected by PhaseNO.
   \item[Figure S16] An example of an M 1.082 event detected by PhaseNO.
   \item[Figure S17] Example of one graph-type sample containing one event from the training dataset.
   \item[Figure S18] Example of one graph-type sample containing two events from the training dataset.
   \item[Figure S19] Learning curves of PhaseNO.
   \item[Figure S20] Prediction errors as a function of SNR, earthquake magnitude, and epicentral distance.
   \item[Figure S21] Predicted probabilities as a function of prediction errors, SNR, earthquake magnitude, and epicentral distance.
   
\end{description}

\setcounter{figure}{0}
\renewcommand{\figurename}{Figure}
\renewcommand{\thefigure}{S\arabic{figure}}

\newpage

\section*{Text S1: Neural Operators}

Neural operators are generalizations of neural networks to map between infinite dimensional function spaces. This new class of models provably satisfy the universal approximation theorem for operators \citep{kovachki_neural_2023}. Here we propose a new architecture for learning maps from wavefields to phase picks. Neural Operators generally begin with a lifting operator ($\mathcal{P}$) that maps the input function $(f)$ to one with a larger co-domain, $v$. These functions are then operated on iteratively with nonlinear kernel integration operators, and finally are passed through a projection operator ($\mathcal{Q}$) that maps the hidden representation to the output function $(g)$. $\mathcal{P}$ and $\mathcal{Q}$ are parameterized with fully connected neural networks and act pointwise on the physical domain. The basic formula of the iterative kernel integration is a composition of linear integral operators and non-linear activation functions. Each integral operator has the following form:

\begin{equation}
u(x) = (\kappa \ast v^l)(x) = \int \kappa(x,y)v^l(y)dy , 
\end{equation}
where $v$ and $u$ are the intermediate input and output functions, respectively, and $\kappa$ is a kernel function. Here, we define $v^1=\mathcal{P}(f)$ as the input to the operator. There are several ways to parameterize the kernel \citep{kovachki_neural_2023}. We treat the seismograms recorded by a seismic network as the input function $f$, discretized with a regular mesh in the time domain and irregular mesh in the spatial domain. In our architecture, we compute the kernel function separately for space and time.

\subsection*{Fourier Neural Operators}

For the regular mesh in the time domain, we parameterize the kernel in Fourier space and compute the kernel integral operator with fast Fourier transform, leading to efficient computation with almost linear complexity \citep{li_fourier_2020}. From the convolution theorem, we have
\begin{equation}
(\kappa \ast v)(x) = \mathcal{F}^{-1}(R_{\phi} \cdot (\mathcal{F}(v))))(x), 
\end{equation}
where $\mathcal{F}$ and $\mathcal{F}^{-1}$ denote the Fourier transform and its inverse. $R_{\phi}$ is the Fourier transform of $\kappa$, parametrized by $\phi$. With an activation function $\sigma$ acting locally in each layer, a single FNO layer update is 

\begin{equation}
u(x) = \sigma ( Wv(x) + \mathcal{F}^{-1}(R_{\phi} \cdot (\mathcal{F}(v)))(x), 
\end{equation}
where $W$ is a local linear operator. In practice, we truncate the Fourier series at a maximal number of modes and parameterize $R$ with a few lower modes. Starting from the input $v$, one FNO layer contains two parallel branches (Figure \ref{fig:model}): one branch computes the kernel in the Fourier space and performs the global integration; the other applies a pointwise linear transform $W$ to the input. Results from two branches are added before applying $\sigma$.

PhaseNO utilizes seven FNO layers similar to the U-NO architecture \citep{rahman2022u}. The number of modes in each FNO layer is 24, 12, 8, 8, 12, 24, and 24. The width (the channel number) of the discretized $u$ at each node changes with the dimension of $R_{\phi}$. At each FNO layer, the discretized $u$ has a dimension of 48×3000, 96×750, 192×200, 96×750, 48×3000, 48×3000, and 48×3000, where the first dimension denotes the width and the second dimension denotes time. All FNO layers include nonlinearity via the Gaussian Error Linear Unit \citep{hendrycks_gaussian_2020}, except the last one where no activation function is applied. Note that we did not draw the last FNO layer on Figure \ref{fig:model} for simplicity.

\subsection*{Graph Neural Operators and the message passing framework}

Kernel integration can be viewed as an aggregation of messages with the generalized message passing in graph neural networks \citep{li_neural_2020}. Since $f(x,y,t)$ in the spatial domain is discretized based on the geometry of a seismic network, we parameterize the kernel with GNOs and implement it with the message passing framework. We consider a seismic network with an arbitrary geometry as a graph. Each station in the seismic network is a node of the graph. Given node features $v(x)$, we update the value $v(x_i)$ of the node $x_i$ to the value $u(x_i)$ with the averaging aggregation by
\begin{equation}
u(x_i) = \tau (  v(x_i), \frac{1}{|\mathcal{N}(x_i)|} \sum_{x_j \in \mathcal{N}(x_i)} \varphi( v(x_i), v(x_j) ) ) , 
\end{equation}
where $\tau$ and $\varphi$ denote differentiable functions such as multilayer perceptron (MLP). $n$ is the number of nodes in a graph. $\mathcal{N}(x_i)$ denotes the neighborhood of $x_i$. To capture global dependencies, we construct a graph by connecting each node to all nodes in the graph, resulting in a total edge number of $n^2$. In other words, $\mathcal{N}(x_i)$ consists of all stations in a seismic network and includes a self-loop, meaning that each node has an edge pointing to itself. 

The edge features are computed by $e_{ij}=\varphi(v(x_i),v(x_j))$ where $\varphi: \mathbb{R}^c \times \mathbb{R}^c \to \mathbb{R}^{c'}$ is a nonlinear function with a set of learnable parameters \citep{wang_dynamic_2019}. We choose $\varphi$ as an MLP with one hidden layer containing $4c$ neurons. The function takes the concatenation of two node features $v(x_i)$ and $v(x_j)$ as the input (with a channel number of $2c$) and outputs $e_{ij}$ with the same channel number as the node features $(c'=c)$. When all the edge features (messages) are available, the target node $x_i$ collects all the messages and aggregates them with an averaging operation. Finally, we use another MLP for $\tau$ (with the same architecture as $\varphi$) to update the nodes features of $x_i$ using the concatenation of $v(x_i)$ and the aggregated message as the input. Message passing allows the exchange of information between neighboring nodes, which enhances the relevant signals shared by adjacent nodes. 

\section*{Text S2: Encoding station locations as node features}

Station locations are encoded as node features along with waveform data. Instead of directly using longitudes and latitudes, here we convert the geographic locations ($a_i$,$b_i$) of stations on the Earth to their relative locations ($x_i$,$y_i$) on the computational domain. The converted locations $x_i$,$y_i$ are included as two channels of the input along with three-component waveforms. Each sample has a computational domain varying in its center. The center is selected based on the maximum longitude ($a_{max}$), the minimum longitude ($a_{min}$), the maximum latitude ($b_{max}$), and the minimum latitude ($b_{min}$) of all stations in a graph. The physical minimum of the computational domain is 
\begin{equation}
    a_0 = \frac{a_{max}+a_{min}}{2} - \frac{l}{2}   
\end{equation}
\begin{equation}
    b_0 = \frac{b_{max}+b_{min}}{2} - \frac{l}{2}   
\end{equation}
where $l$ denotes the physical range of the computational domain on the Earth. Then the relative location of each station on the computational domain is 
\begin{equation}
    x_i = \frac{a_{i}-a_{0}}{l}   
\end{equation}
\begin{equation}
    y_i = \frac{b_{i}-b_{0}}{l}
\end{equation}
The physical range ought to be large enough to include all the selected stations in a graph. Here we choose $l=2^{\circ}$, corresponding to an area of around 200 $km$ $\times$ 200 $km$. The computational domain and the relative locations are calculated independently for each sample during the training process. For practical applications, the locations are determined in the same way but only once if given one seismic network.

\section*{Text S3: Baselines for model evaluation}

In this study, we benchmark the performance of PhaseNO against three leading baseline models: EQTransformer \citep{mousavi_earthquake_2020}, PhaseNet \citep{zhu_phasenet_2018}, and EdgePhase \citep{feng_edgephase_2022}. We summarize key attributes about these baselines here. EQTransformer and PhaseNet are single-station detection and picking models using convolutional layers and other modern deep learning components. PhaseNet was trained on an earthquake dataset from Northern California with several hundred thousand data samples (623,054 P/S picks). EQTransformer was trained on a global dataset of earthquakes called STEAD \citep{mousavi_stanford_2019}. EdgePhase is a multi-station picking model that incorporates an edge convolution module in the latent space of EQTransformer; it was built on the pre-trained layers of EQTransformer and then fine-tuned on earthquake and noise data of the year 2021 recorded by the Southern California Seismic Network (SCSN). The pre-trained EQTransformer compared here has been fine-tuned with the same dataset as EdgePhase, leading to slightly better performance than the original model \citep{feng_edgephase_2022}. Since phase neural pickers usually transfer well between different regions for local earthquakes \citep{munchmeyer_which_2022,zhang2022loc}, it is common to use pre-trained models as baselines for this type of problems \citep{mousavi_earthquake_2020}.

\section*{Text S4: Training and test datasets}

Advanced deep-learning model architecture needs a training dataset with sufficient quality and quantity to investigate its full potential. Taking the wavefield properties of a seismic network into account, we come up with effective data augmentation strategies specifically for graph-type samples. First, we stack events at the graph level aiming to reserve the moveout patterns of arrivals at different stations for each event. In this way, waveforms at different stations in a graph may consist of different number of events. Second, earthquakes follow a power law whereby most events are small and may be observed by only several stations instead of the entire seismic network. Thus, it is important to add virtual stations at random locations in a graph with noise waveforms to regularize PhaseNO.  

We construct a training dataset with three-component earthquake waveforms from NCEDC of the years from 1984 to 2019 and three-component noise waveforms from the STanford EArthquake Dataset (STEAD). The earthquake data are downloaded event by event with stations containing both P and S arrival times picked by human analysts. Gaps are padded with zeros if some segments are missing. We normalize each component by removing its mean and dividing it by the standard deviation. We use a probability function with a triangular shape to label phase arrivals. Probabilities at manually picked P/S arrivals are labeled 1 and linearly decrease to 0 before and after the manual picks. For each pick, the duration of probabilities larger than 0 is 0.4 s, with the highest probability centered on the middle of the time window. Instead of treating seismograms on a single station as one sample, we construct a graph with all stations in a seismic network and use the graph as one sample. 

We perform data augmentation during training. We stack the individually downloaded events with the following steps to reserve their moveout patterns on different stations:
\begin{itemize}
\item Randomly select station A from all stations recording event A, 
\item Randomly select event B from all the events recorded by station A,
\item Randomly assign the weights of $\alpha$ and $\beta$ ($0.1 < \alpha < 0.9$, $0.1<\beta< 0.9$, $\alpha+\beta$=1) to the amplitudes of two events,
\item Randomly select a time shift between two events to be stacked,
\item Stack event A and event B if both events are recorded on the same stations, and
\item Reserve waveforms at other stations that record only one event. 
\end{itemize}
Generally, more than one station records both events in a seismic network, even though we select event B based on one station recording event A. More events can be stacked by repeating the above steps. Around 66$\%$ of samples in the training dataset contain two or three events. We also generate up to 16 virtual stations with random locations and assign noise waveforms to these virtual stations. The noise data are randomly selected from the 235K noise samples in the STEAD. Except for 6.25$\%$ of samples that contain only earthquake waveforms, each sample has both earthquake and noise waveforms recorded at different stations in a seismic network. In one graph-type sample, the number of events on each station ranges from zero to three. Since a seismic network may contain both three-component and one-component seismometers, we randomly select several stations and consider them as one-component stations. On these stations, we randomly select and repeat one component three times. Each sample may have different number of stations. To save the computational cost, we reserve no more than 32 but at least 5 stations in one graph. We then cut 30-s waveforms at all stations with a random starting time, so positions of phases within the window are varied. With a sampling rate of 100 Hz, both input waveform and output probability have 3000 data points for each component at each station. In total we have 57K graphs for training. The edge index is built during training, with all nodes in a graph. If one station contains multiple types of channels, we consider them as different nodes with the same geographic locations. We show two examples of the graph-type samples in the Supporting Information (Figures \ref{figs17:graph_example1} and \ref{figs18:graph_example2}).

The input data at each station consist of five channels: three waveform channels and two location channels. The output has two channels with P-phase and S-phase probabilities. To encode station locations as node features, we define a computational domain of $2^{\circ} \times 2^{\circ}$ in the longitude and latitude dimensions and convert the geographic locations of stations to their relative locations on the computational domain. The converted locations $x_i$, $y_i$ are included as the fourth and fifth channels along with waveforms.

The test dataset contains 5,769 samples built with the NCEDC earthquake dataset of the year 2020. Waveforms are preprocessed in a similar way to the training dataset without data augmentation. Each sample in the test dataset contains only one event recorded by multiple stations. In total, we use 43,700 P/S picks of the 5,769 events to evaluate PhaseNO and compare with other methods.

\section*{Text S5: Training details and evaluation metrics}

We adopt the binary cross-entropy as our loss function. We choose Adaptive Moment Estimation (Adam) as the optimizer with a batch size of one and a learning rate of $10e^{-4}$. The training takes around 10 hours for each epoch on one NVIDIA Tesla V100 GPU. The model converges after around 10 epochs (Figure \ref{figs19:loss}). We stop training after 20 epochs and use the result as our final model. 

We use six metrics to evaluate the performance: precision, recall, F1 score, mean ($\mu$), standard deviation ($\sigma$), and mean absolute error (MAE) between resulting picks and human analysts (labels). The resulting pick is counted as a true positive pick (TP) if the time residual between the pick and the label is less than 0.5 s. If there are no positive picks beyond a time residual of 0.5 s for one label, we count it as a false negative one (FN). Moreover, false positives are counted if the positive picks cannot match any label (FP). Then we can evaluate the performance with:
\begin{equation}
    Precision = \frac{TP}{TP+FP}
\end{equation}
\begin{equation}
    Recall = \frac{TP}{TP+FN}
\end{equation}
\begin{equation}
    F1 score = \frac{2 \times Precision \times Recall}{Precision+Recall}
\end{equation}

\section*{Text S6: More discussions}

By analyzing the prediction errors as a function of SNR, earthquake magnitude, and epicentral distance (Figure \ref{figs15:factors_errors}), we showed that SNR plays a more important role than the other factors. Errors tend to increase with decreasing SNR, but they are generally consistent at different magnitudes. Large epicentral distances do not necessarily cause large prediction errors, probably because earthquakes that can propagate to long distances are generally large ones and show high SNR signals over a wide area. We further examine the relationship between these factors with predicted probabilities (Figure \ref{figs16:factors_probs}). We find that low prediction probabilities generally appear at low SNRs. Other factors, such as earthquake magnitude and epicentral distance, seem to have slight impact on the predicted probabilities. Moreover, most phases can be accurately detected with minor prediction errors, although the prediction probabilities may be low. More confident predictions unexpectedly show larger prediction errors than less confident predictions (smaller probabilities), which may imply the imperfection of manually picked labels. 

We found that the uncertainty in PhaseNO predictions may correlate with the peak prediction width. However, wider peaks on low fidelity signals may impose challenges on pick determination, resulting in slightly increased errors of arrival times if just the maximum value is saved, particularly for S phases (Figure \ref{figs10:ridgecrest_picks}). These uncertainties would thus be of greater value for location or tomography algorithms that consider measurement errors explicitly. 

\newpage

\begin{table}[]
\renewcommand\thetable{S1}
\caption{\label{tab:scores}Phase picking performance on the test dataset. The best score is highlighted in bold.}
\centering
\begin{tabular}{clcccccc}
\hline
\multicolumn{2}{}{}                      & $\mu$(s)  & $\sigma$(s)  & MAE(s) & F1   & Precision & Recall \\ \hline
\multirow{4}{*}{P-phase} & PhaseNO       & \textbf{0.00} & 0.05 & \textbf{0.02}   & \textbf{0.99} & \textbf{0.99}      & \textbf{0.99}   \\
                         & PhaseNet      & \textbf{0.00} & \textbf{0.04} & \textbf{0.02}   & 0.96 & 0.95      & 0.97   \\
                         & EdgePhase     & 0.01 & 0.09 & 0.07   & 0.97 & 0.98      & 0.96   \\
                         & EQTransformer & 0.02 & 0.08 & 0.06   & 0.96 & 0.97      & 0.95   \\ \hline
\multirow{4}{*}{S-phase} & PhaseNO       & \textbf{0.01} & \textbf{0.09} & \textbf{0.06}   & \textbf{0.98} & \textbf{0.97}      & \textbf{0.98}   \\
                         & PhaseNet      & 0.02 & \textbf{0.09} & \textbf{0.06}   & 0.93 & 0.90      & 0.96   \\
                         & EdgePhase     & 0.07 & 0.12 & 0.11   & 0.95 & 0.96      & 0.95   \\
                         & EQTransformer & \textbf{0.01} & 0.12 & 0.09   & 0.94 & 0.95      & 0.93   \\ \hline
\end{tabular}
\end{table}

\newpage

\begin{table}[]
\renewcommand\thetable{S2}
\centering
\caption{List of stations used for PhaseNO and PhaseNet in the 2019 Ridgecrest earthquake sequence.}
\label{tab:stations}
\begin{tabular}{ccccccc}
\hline
   & network & station & channel & latitude & longitude & component \\ \hline
1  & CI      & CCC     & HH      & 35.525   & -117.365  & E,N,Z     \\
2  & CI      & CCC     & HN      & 35.525   & -117.365  & E,N,Z     \\
3  & CI      & CLC     & HH      & 35.816   & -117.598  & E,N,Z     \\
4  & CI      & CLC     & HN      & 35.816   & -117.598  & E,N,Z     \\
5  & CI      & DAW     & HH      & 36.271   & -117.592  & E,N,Z     \\
6  & CI      & DAW     & HN      & 36.271   & -117.592  & E,N,Z     \\
7  & CI      & JRC2    & HH      & 35.982   & -117.809  & E,N,Z     \\
8  & CI      & JRC2    & HN      & 35.982   & -117.809  & E,N,Z     \\
9  & CI      & LRL     & HH      & 35.48    & -117.682  & E,N,Z     \\
10 & CI      & LRL     & HN      & 35.48    & -117.682  & E,N,Z     \\
11 & CI      & MPM     & HH      & 36.058   & -117.489  & E,N,Z     \\
12 & CI      & MPM     & HN      & 36.058   & -117.489  & E,N,Z     \\
13 & CI      & SLA     & HH      & 35.891   & -117.283  & E,N,Z     \\
14 & CI      & SLA     & HN      & 35.891   & -117.283  & E,N,Z     \\
15 & CI      & SRT     & HH      & 35.692   & -117.751  & E,N,Z     \\
16 & CI      & SRT     & HN      & 35.692   & -117.751  & E,N,Z     \\
17 & CI      & TOW2    & HH      & 35.809   & -117.765  & E,N,Z     \\
18 & CI      & TOW2    & HN      & 35.809   & -117.765  & E,N,Z     \\
19 & CI      & WBM     & HH      & 35.608   & -117.89   & E,N,Z     \\
20 & CI      & WBM     & HN      & 35.608   & -117.89   & E,N,Z     \\
21 & CI      & WCS2    & HH      & 36.025   & -117.765  & E,N,Z     \\
22 & CI      & WCS2    & HN      & 36.025   & -117.765  & E,N,Z     \\
23 & CI      & WMF     & HH      & 36.118   & -117.855  & E,N,Z     \\
24 & CI      & WMF     & HN      & 36.118   & -117.855  & E,N,Z     \\
25 & CI      & WNM     & EH      & 35.842   & -117.906  & Z         \\
26 & CI      & WNM     & HN      & 35.842   & -117.906  & E,N,Z     \\
27 & CI      & WRC2    & HH      & 35.948   & -117.65   & E,N,Z     \\
28 & CI      & WRC2    & HN      & 35.948   & -117.65   & E,N,Z     \\
29 & CI      & WRV2    & EH      & 36.008   & -117.89   & Z         \\
30 & CI      & WRV2    & HN      & 36.008   & -117.89   & E,N,Z     \\
31 & CI      & WVP2    & EH      & 35.949   & -117.818  & Z         \\
32 & CI      & WVP2    & HN      & 35.949   & -117.818  & E,N,Z     \\
33 & PB      & B916    & EH      & 36.193   & -117.668  & 1,2,Z     \\
34 & PB      & B917    & EH      & 35.405   & -117.259  & 1,2,Z     \\
35 & PB      & B918    & EH      & 35.936   & -117.602  & 1,2,Z     \\
36 & PB      & B921    & EH      & 35.587   & -117.462  & 1,2,Z     \\ \hline
\end{tabular}
\end{table}

\newpage

\begin{table}[]
\renewcommand\thetable{S3}
\centering
\caption{Comparison of association results for picks detected by PhaseNO and PhaseNet in the 2019 Ridgecrest earthquake sequence.}
\label{tab:association}
\begin{tabular}{ccccccc}
\hline
Method                    & \begin{tabular}{@{}c@{}c@{}}Phase\\picking\\threshold\end{tabular}
 & \begin{tabular}{@{}c@{}}Total\\picks\end{tabular}
                & \begin{tabular}{@{}c@{}c@{}}Minimum\\picks in\\GaMMA\end{tabular} & \begin{tabular}{@{}c@{}c@{}}Total\\associated\\events\end{tabular} & \begin{tabular}{@{}c@{}c@{}}Total\\associated\\picks\end{tabular} &  \begin{tabular}{@{}c@{}c@{}}Number of\\picks per\\event\end{tabular} \\ \hline
\multirow{5}{*}{PhaseNO}  & \multirow{3}{*}{0.3}    & \multirow{3}{*}{1,379,895} & 12                     & 33,142                  & 1,126,949              & 34.00                     \\
                          &                         &                            & 15                     & 28,624                  & 1,068,563              & 37.33                     \\
                          &                         &                            & 17                     & 26,176                  & 1,030,665              & 39.37                     \\ \cline{2-7} 
                          & \multirow{2}{*}{0.4}    & \multirow{2}{*}{1,184,681} & 12                     & 28,840                  & 976,929                & 33.87                     \\
                          &                         &                            & 15                     & 25,039                  & 927,736                & 37.05                     \\ \hline
\multirow{5}{*}{PhaseNet} & \multirow{3}{*}{0.3}    & \multirow{3}{*}{1,115,784} & 12                     & 28,099                  & 904,228                & 32.18                     \\
                          &                         &                            & 15                     & 23,965                  & 850,807                & 35.50                     \\
                          &                         &                            & 17                     & 21,748                  & 816,485                & 37.54                     \\ \cline{2-7} 
                          & \multirow{2}{*}{0.4}    & \multirow{2}{*}{1,026,533} & 12                     & 26,176                  & 835,257                & 31.91                     \\
                          &                         &                            & 15                     & 22,369                  & 786,025                & 35.14                     \\ \hline
\end{tabular}
\end{table}

\begin{table}[]
\renewcommand\thetable{S4}
\centering
\caption{Comparison of training datasets.}
\label{tab:dataset}
\begin{tabular}{cccc}
\hline
Method        & Size   & Region  & Reference  \\ \hline
PhaseNO       & \begin{tabular}{@{}l@{}l@{}l@{}l@{}}57,000 graph-type samples.\\About 93.75$\%$ of samples\\contain both earthquake and\\noise waveforms. Others are\\only earthquake samples\end{tabular}     & \begin{tabular}{@{}l@{}l@{}}Earthquake from\\NCEDC (1984-2019);\\Noise from STEAD\end{tabular} & This study \\ \hline
PhaseNet      & 623,054 picks   & NCEDC (1984-2019)   & Zhu and Beroza, 2019 \\ \hline
EdgePhase     & \begin{tabular}{@{}l@{}l@{}}12,718 graph-type earthquake\\samples + 15,813 graph-type\\noise samples\end{tabular}  & \begin{tabular}{@{}l@{}l@{}l@{}}Pretrained with\\STEAD (global);\\Finetuned with\\SCSN (2021)\end{tabular} & Feng et al., 2022    \\ \hline
EQTransformer & \begin{tabular}{@{}l@{}}1M earthquake waveforms\\and 300K noise waveforms\end{tabular}   & \begin{tabular}{@{}l@{}l@{}l@{}}Pretrained with\\STEAD (global);\\Finetuned with\\SCSN (2021)\end{tabular} & Mousavi et al., 2020 \\ \hline
\end{tabular}
\end{table}

\newpage

\begin{figure}
\centering
\includegraphics[width=0.8\textwidth]{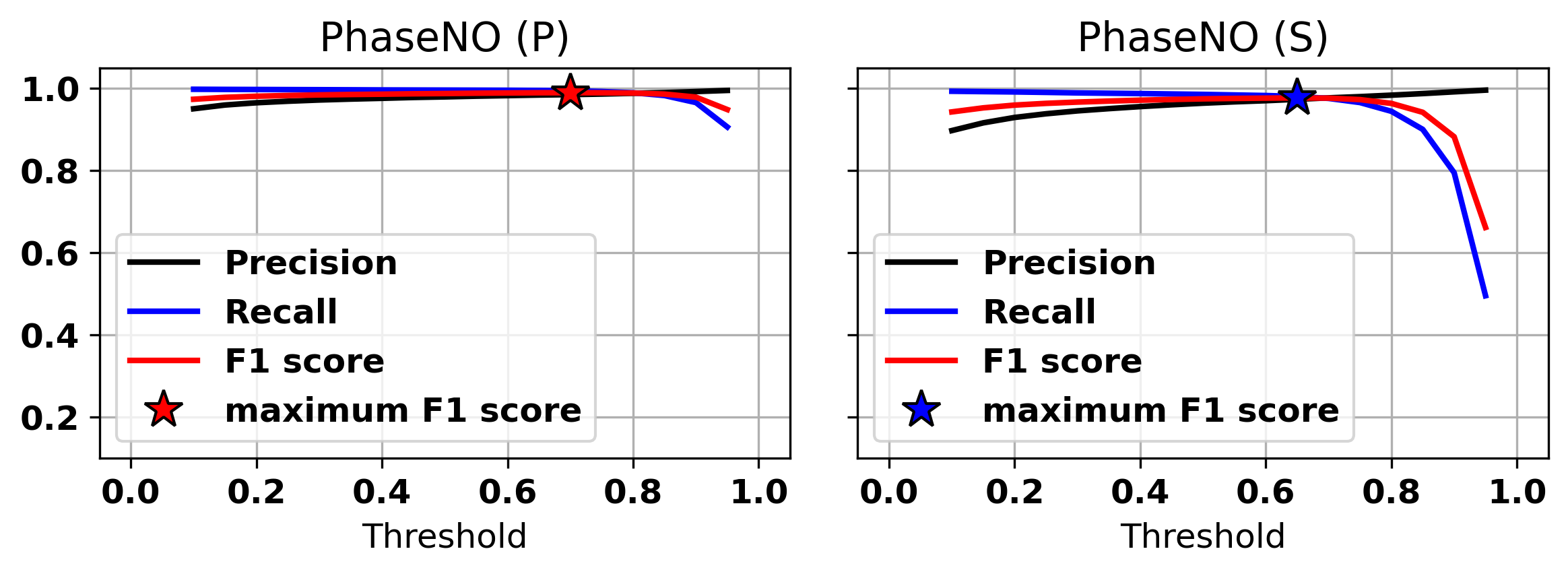}
\includegraphics[width=0.8\textwidth]{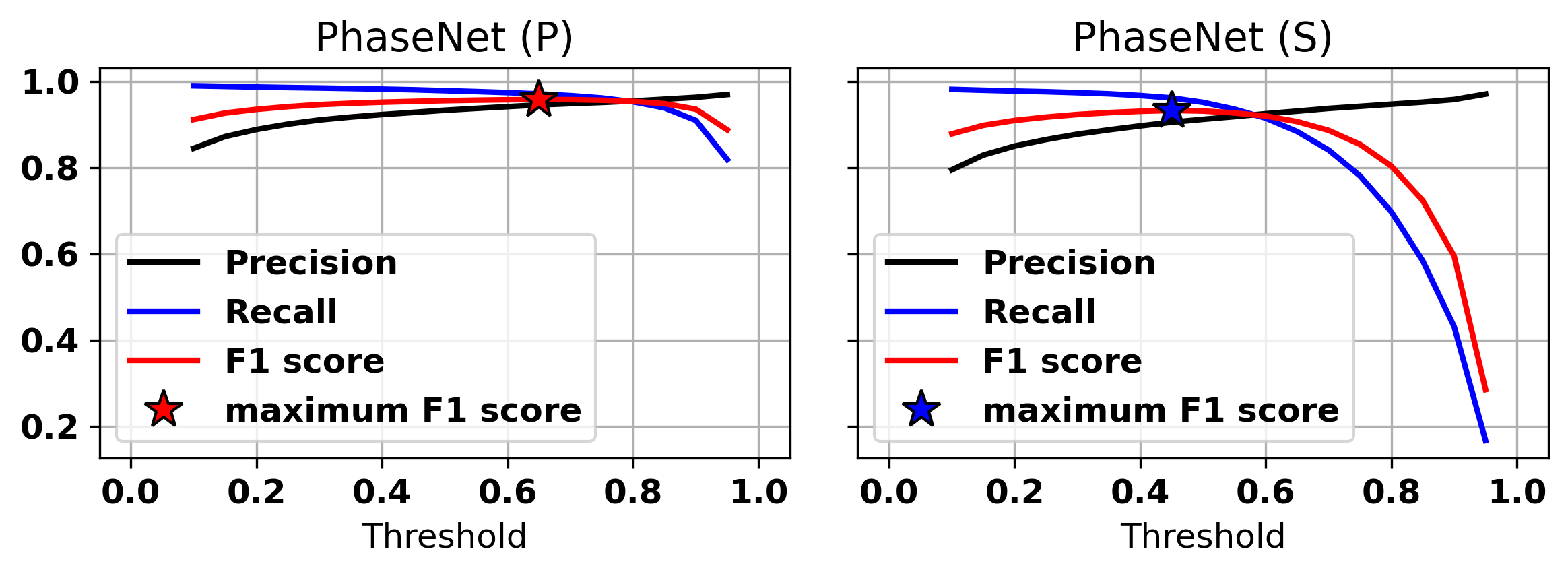}
\includegraphics[width=0.8\textwidth]{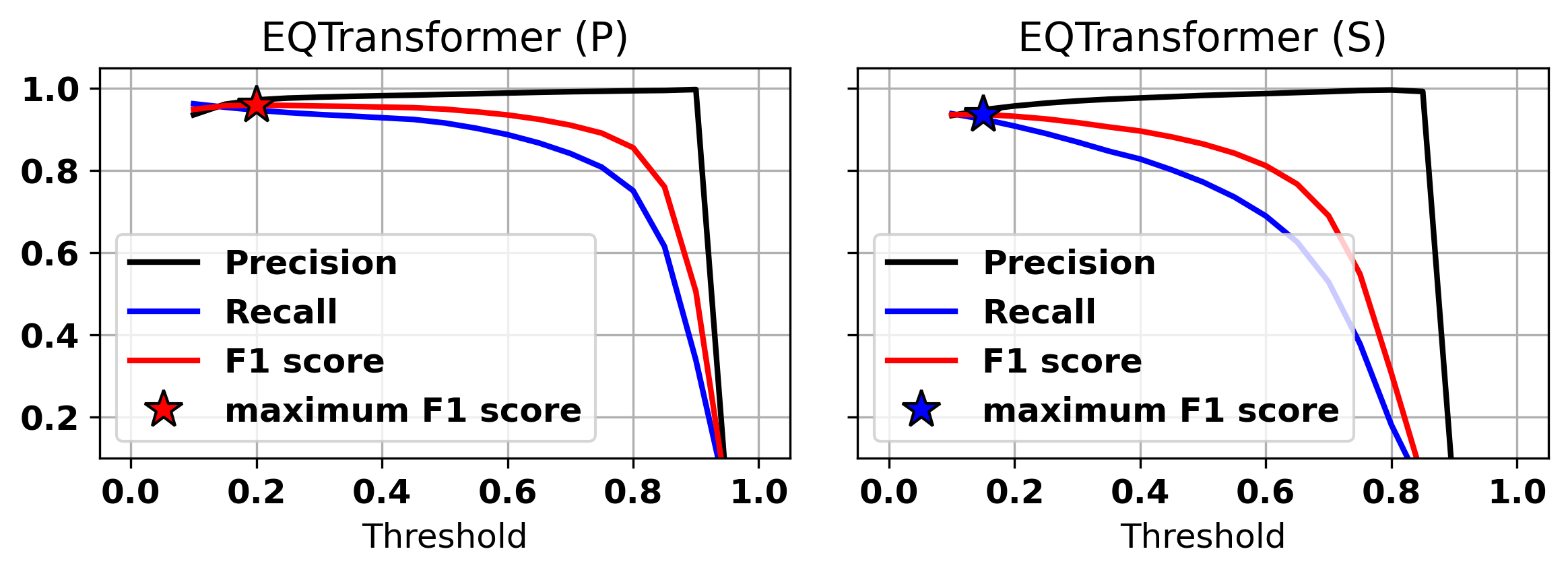}
\includegraphics[width=0.8\textwidth]{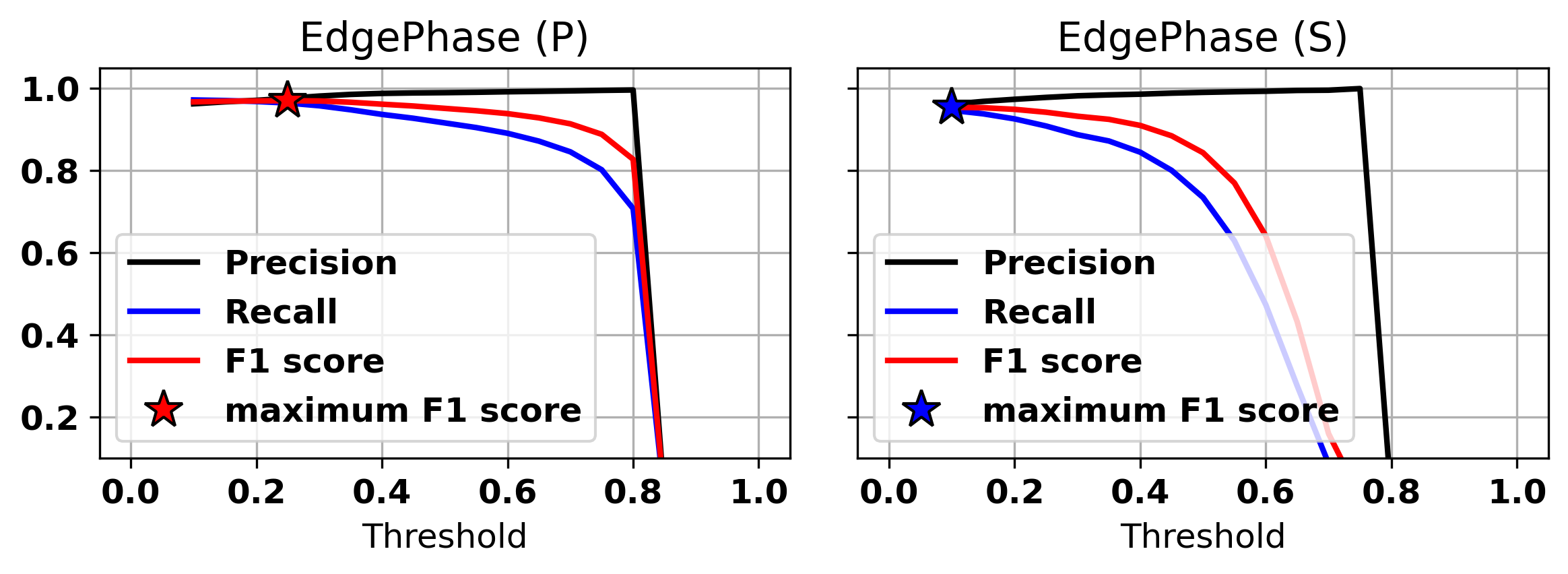}
\caption{\textbf{Comparison of the threshold sensitivity for PhaseNO, PhaseNet, EQTransformer, and EdgePhase on the test dataset.} We select the threshold of each model based on the maximum F1 scores that models can achieve (stars labeled on curves).} 
\label{figs1:threshold}
\end{figure}

\begin{figure}
\centering
\includegraphics[width=0.8\textwidth]{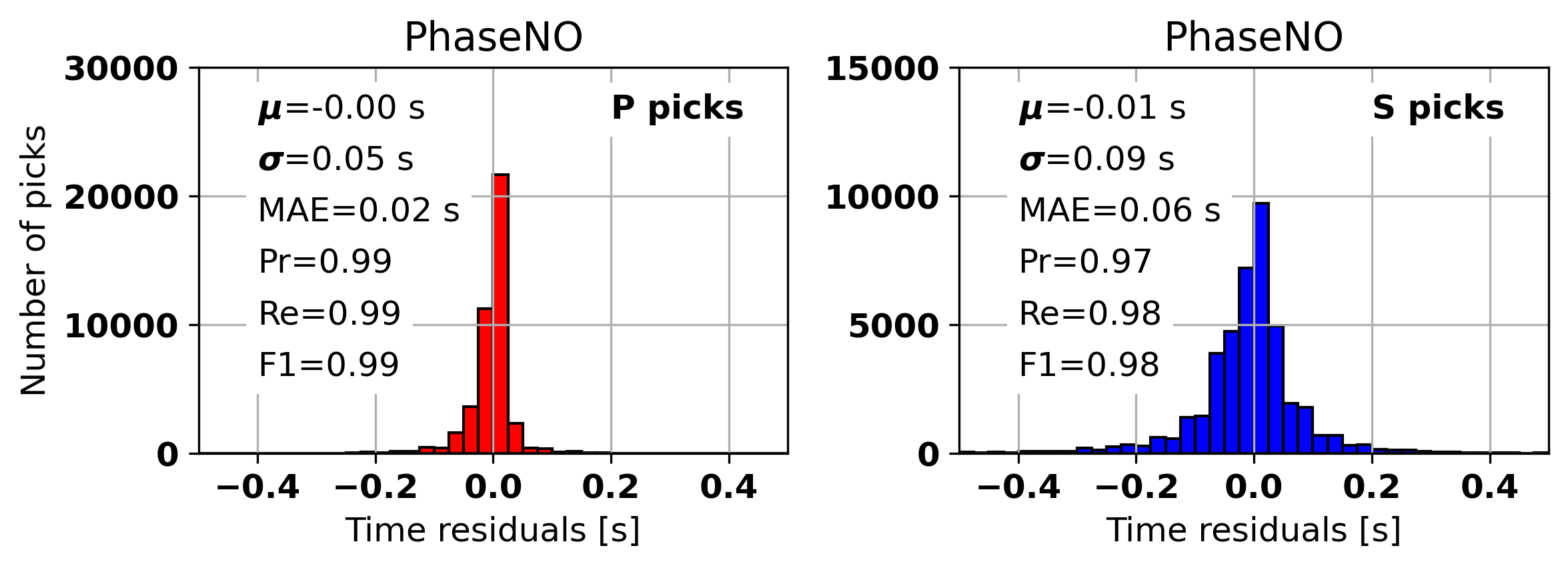}
\includegraphics[width=0.8\textwidth]{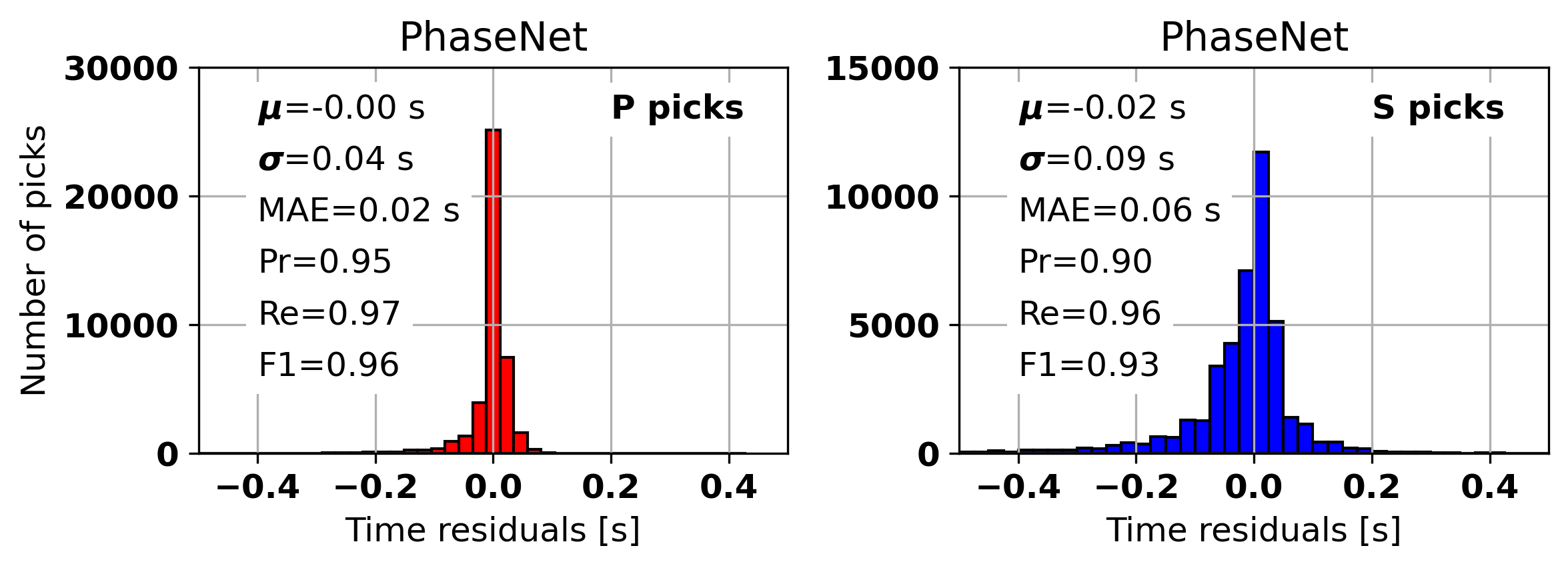}
\includegraphics[width=0.8\textwidth]{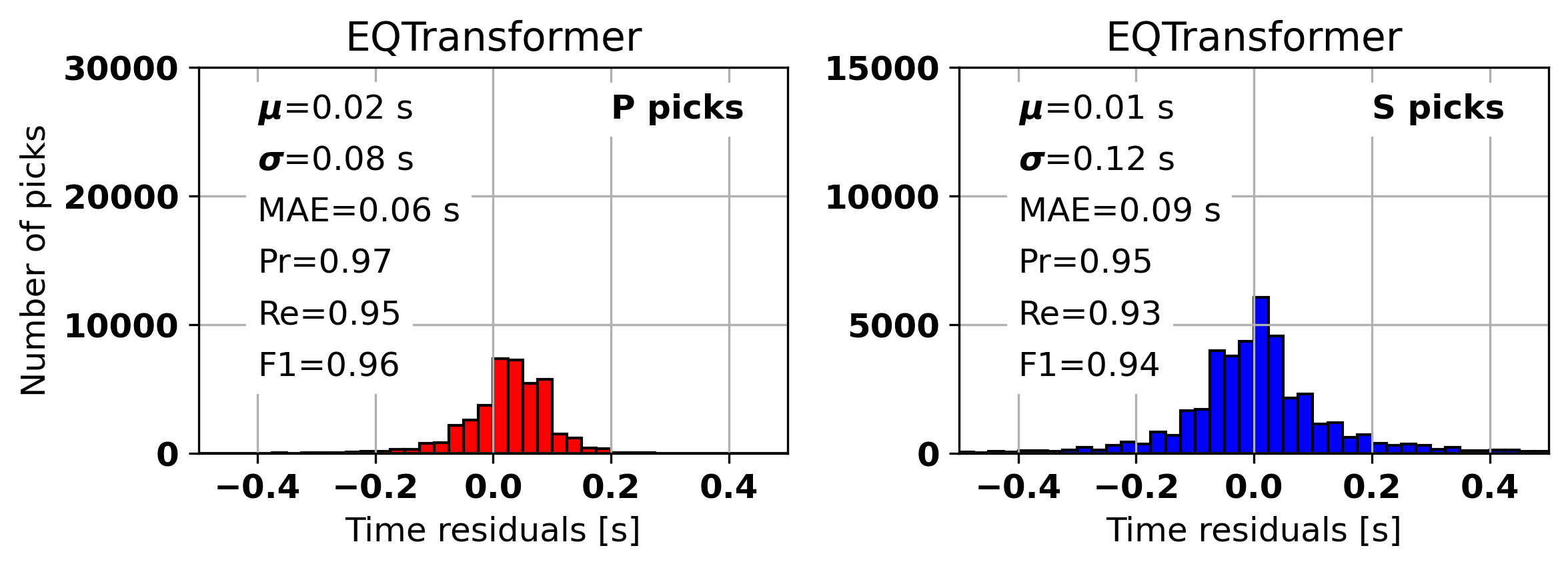}
\includegraphics[width=0.8\textwidth]{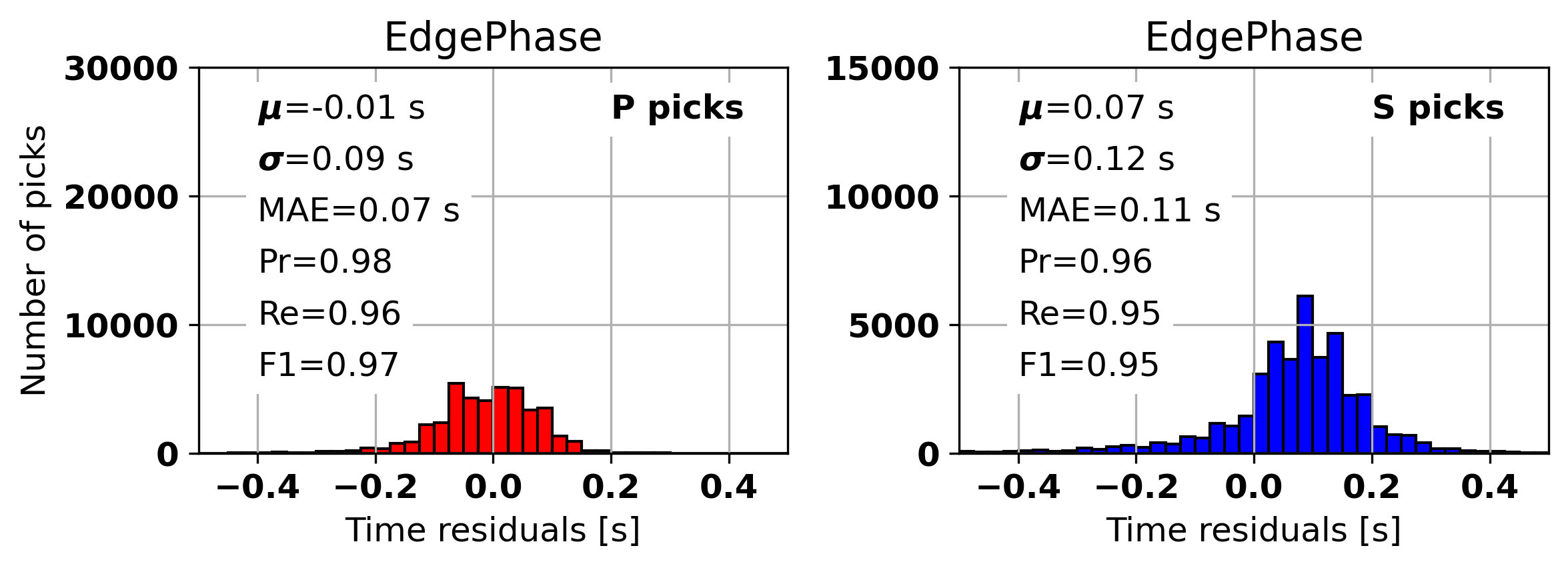}
\caption{\textbf{Comparison of picking errors of deep-learning models on the test dataset.} Time residuals are the travel time difference between machine-learning pickers and manual labels. Mean ($\mu$), standard deviation ($\sigma$) and mean absolute value (MAE) are computed with all true positive picks, which are considered when the time residual is less than 0.5 s compared with manual labels. In addition, precision (Pr), recall (Re) and F1-scores are shown on each panel.} 
\label{figs2:hist}
\end{figure}

\begin{figure}
\centering
\includegraphics[width=\textwidth]{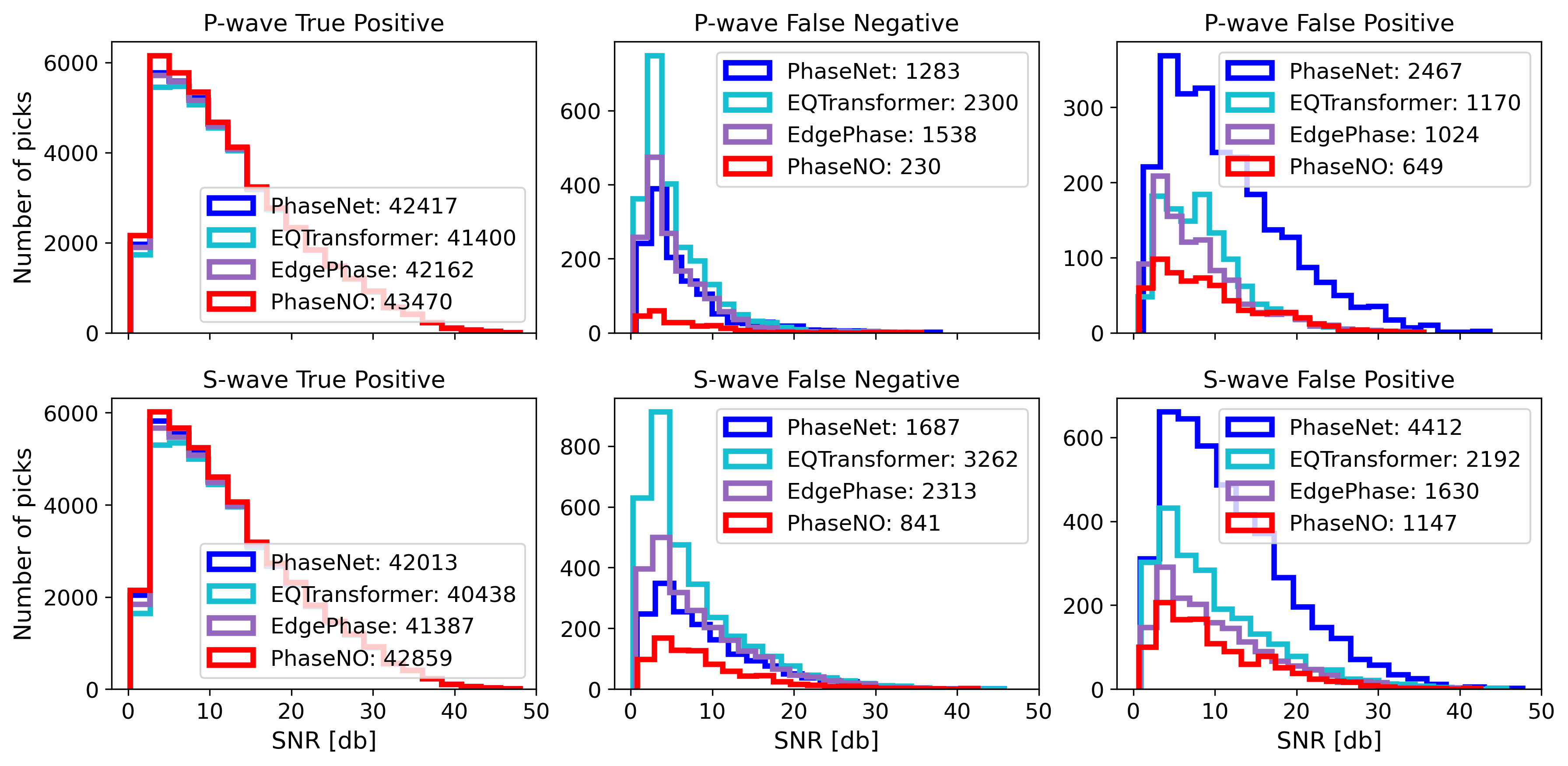}
\caption{\textbf{Phase picking performance as a function of noise level.} PhaseNO detects the most P and S phases and the fewest false picks compared to other state-of-the-art deep-learning models at almost all signal-to-noise ratio (SNR) levels. SNR is calculated by the ratio of standard deviations of the 5 s following and the 5 s preceding the arrival time of P phases.}
\label{fig:snr_histogram}
\end{figure}

\begin{figure}
\centering 
\subfloat[]{\label{a}\includegraphics[width=\textwidth]{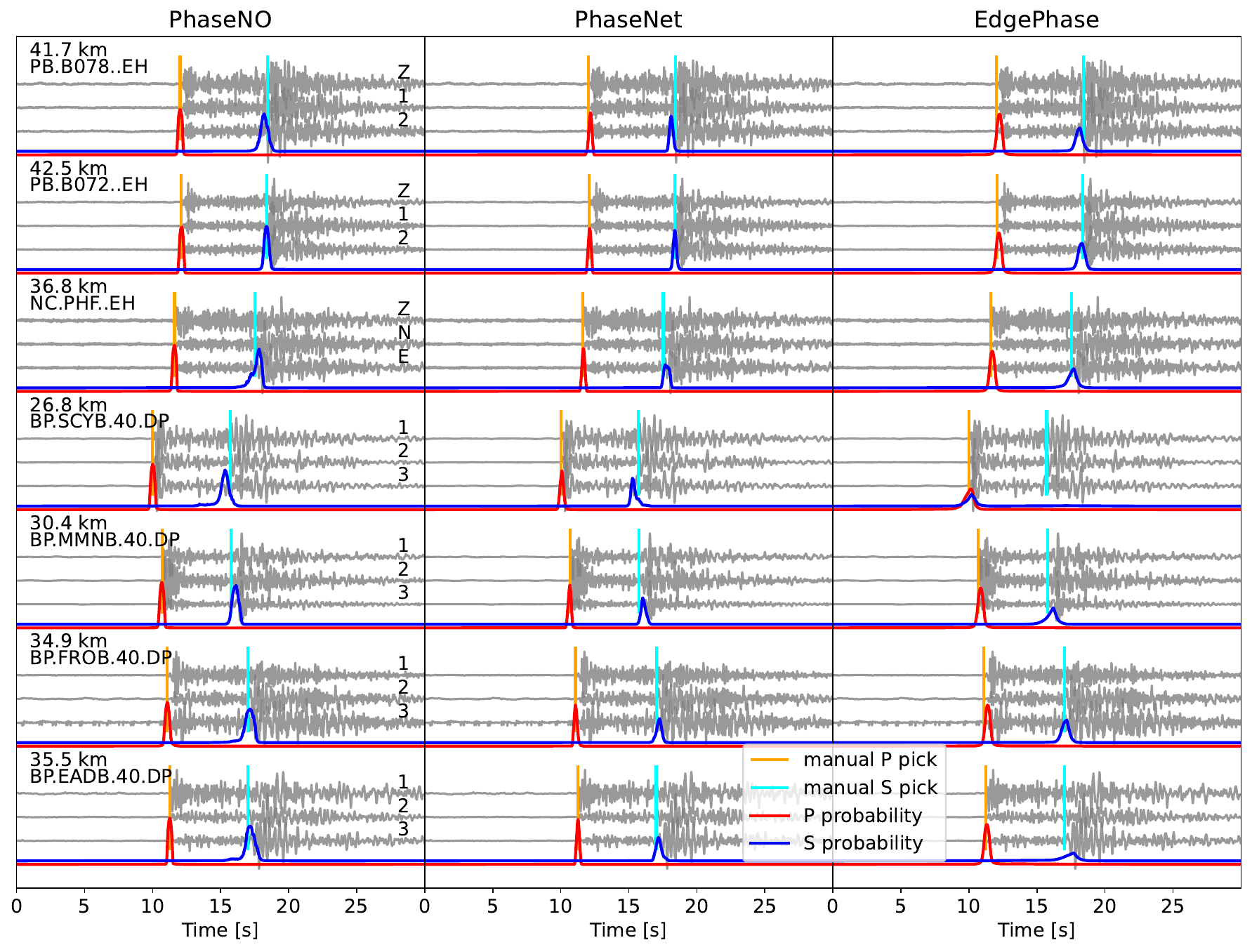}} 
\\
\subfloat[]{\label{b}\includegraphics[width=\textwidth]{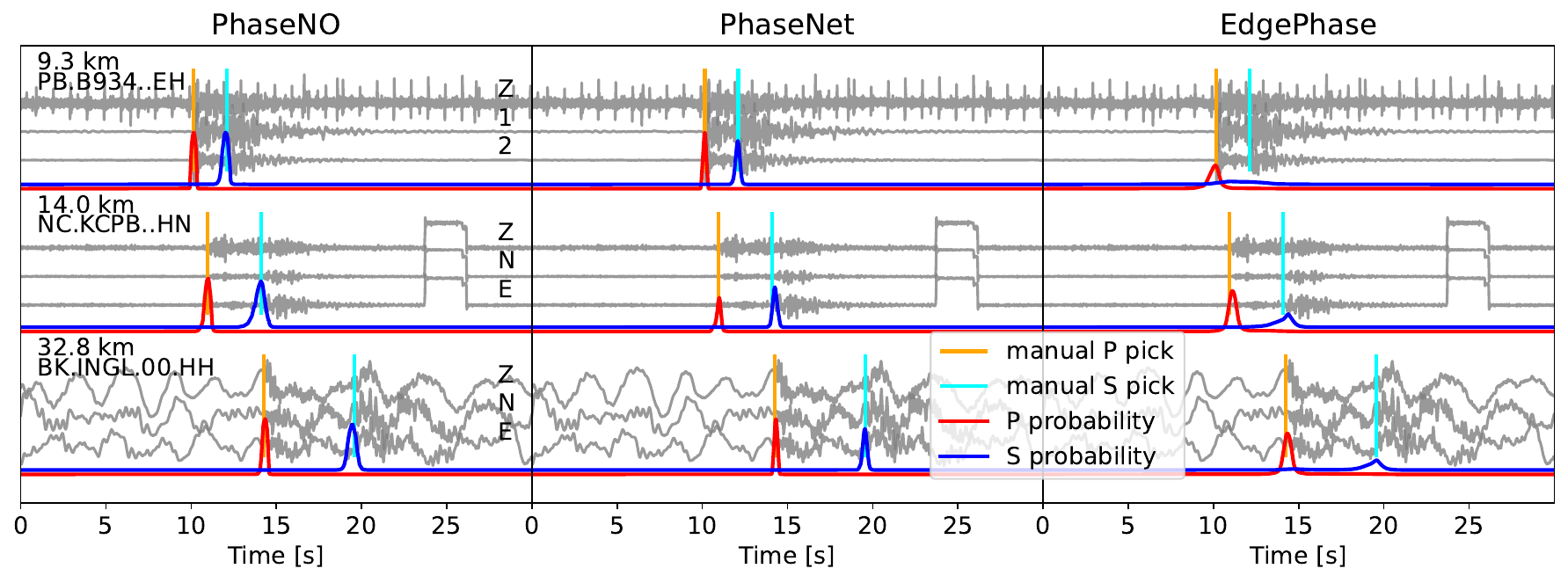}} 
\caption{\textbf{Results of two representative events in the test dataset.} (a) Event nc71112269 with the magnitude of 2.04. The manual labels of the S pick are probably wrong on several stations of this event. (b) Event nc71112684 with the magnitude of 1.41. The station name and epicentral distance are shown on the left part of the three-component waveforms. PhaseNO and EdgePhase predict the results event by event, whereas PhaseNet outputs the results station by station.} 
\label{figs4:test_waveforms}
\end{figure}

\begin{figure}
\centering
\includegraphics[width=\textwidth]{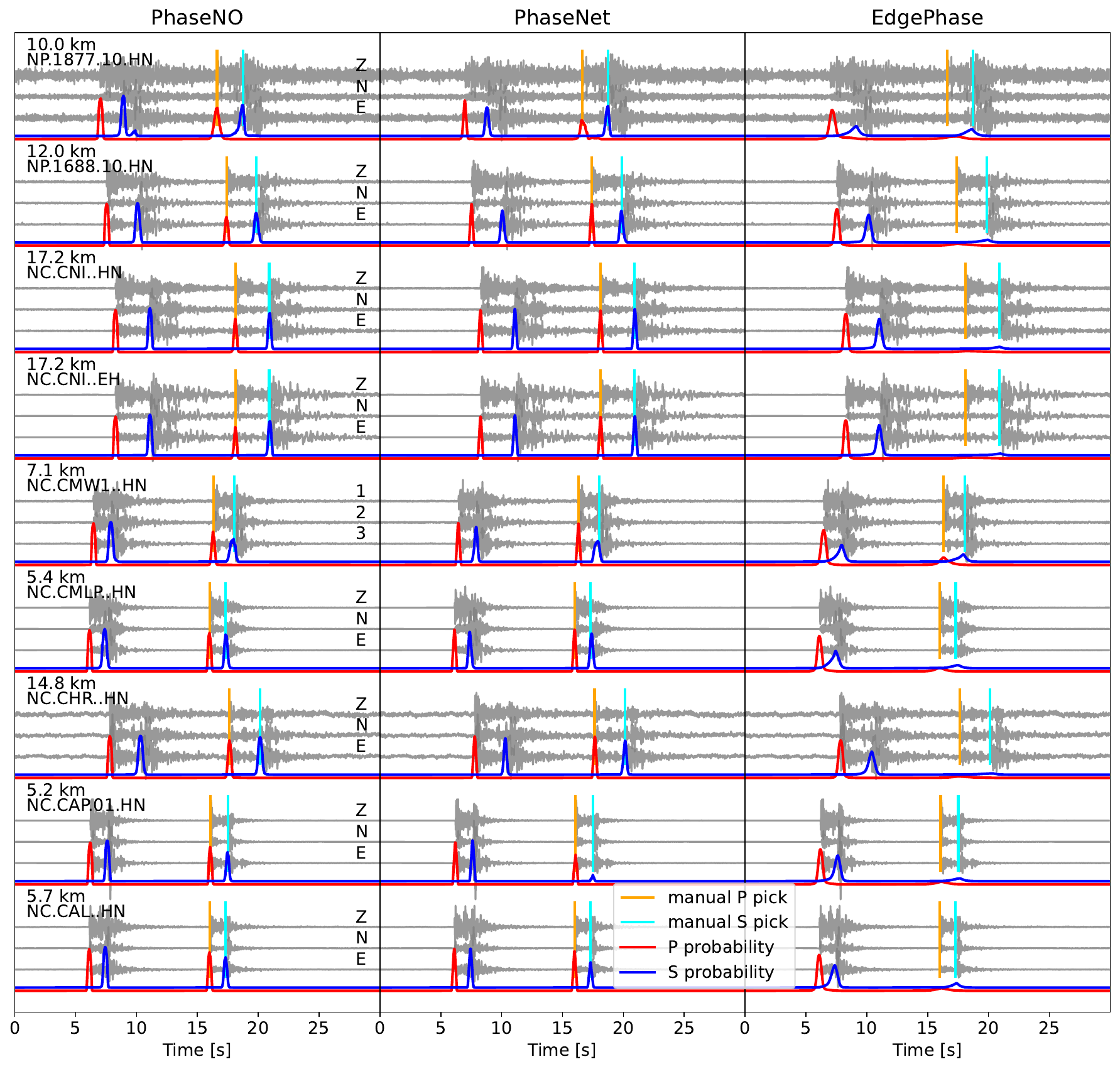}
\caption{\textbf{More examples from the test dataset.} This sample contains two events but only the second one with the magnitude of 1.96 has manual labels.} 
\label{figs5:test_waveforms}
\end{figure}

\begin{figure}
\centering
\includegraphics[width=0.8\textwidth]{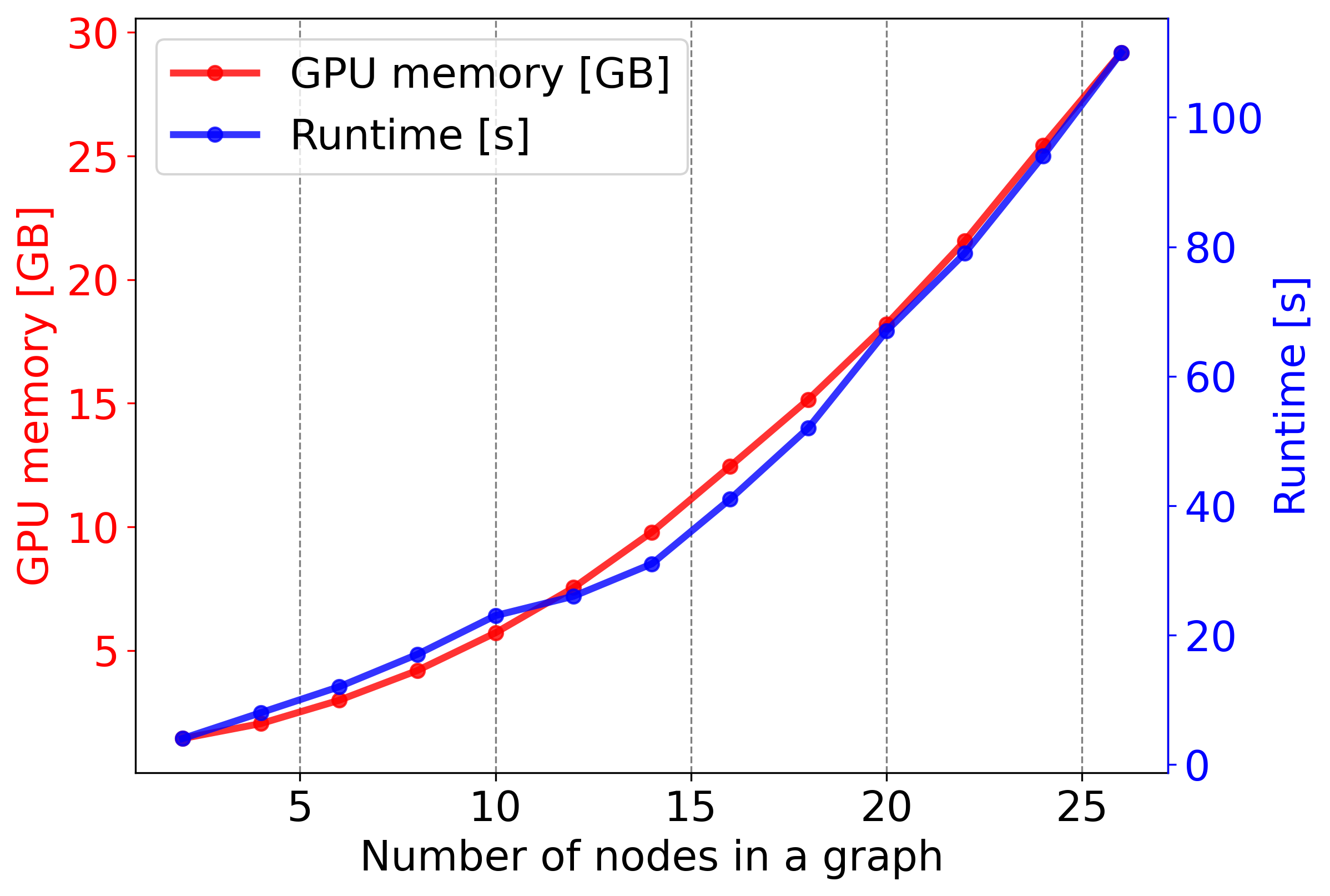}
\caption{\textbf{PhaseNO scaling properties as a function of the number of nodes in one graph}: memory usage and computational time for processing one-hour continuous data that were cut into 180 30-s time windows with an overlap of 10 s. The runtime of PhaseNO scales linearly with the duration of continuous waveforms, and the memory usage remains constant for a fixed number of stations in a seismic network. The offline training stage is performed only once and thus excluded from the runtime evaluation. }
\label{figs5:cost}
\end{figure}

\begin{figure}
\centering
\includegraphics[width=\textwidth]{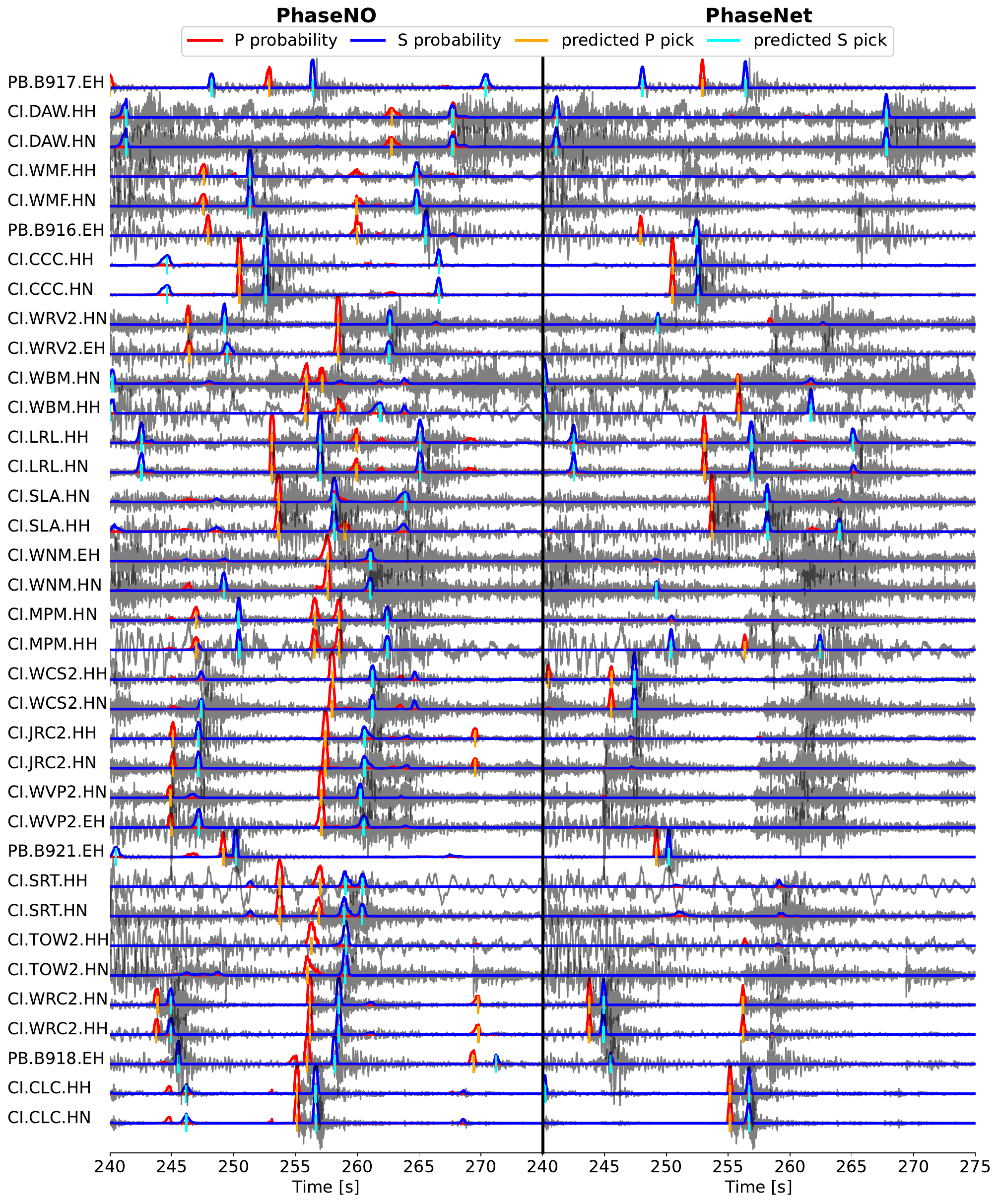}
\caption{\textbf{Representative waveforms selected from one-hour continuous data starting at 06:00:00 on July 6, 2019 from the 2019 Ridgecrest earthquake sequence.} We compare PhaseNO and PhaseNet on a 35-s time window (240 – 275 s). The picking threshold is 0.3. PhaseNO successfully detects phases on various shapes of waveforms with many more picks than PhaseNet. We notice that most of the newly detected phases are meaningful by visually checking the waveforms.} 
\label{fig:ridgecrest_waveform_070606_240}
\end{figure}

\begin{figure}
\centering
\includegraphics[width=\textwidth]{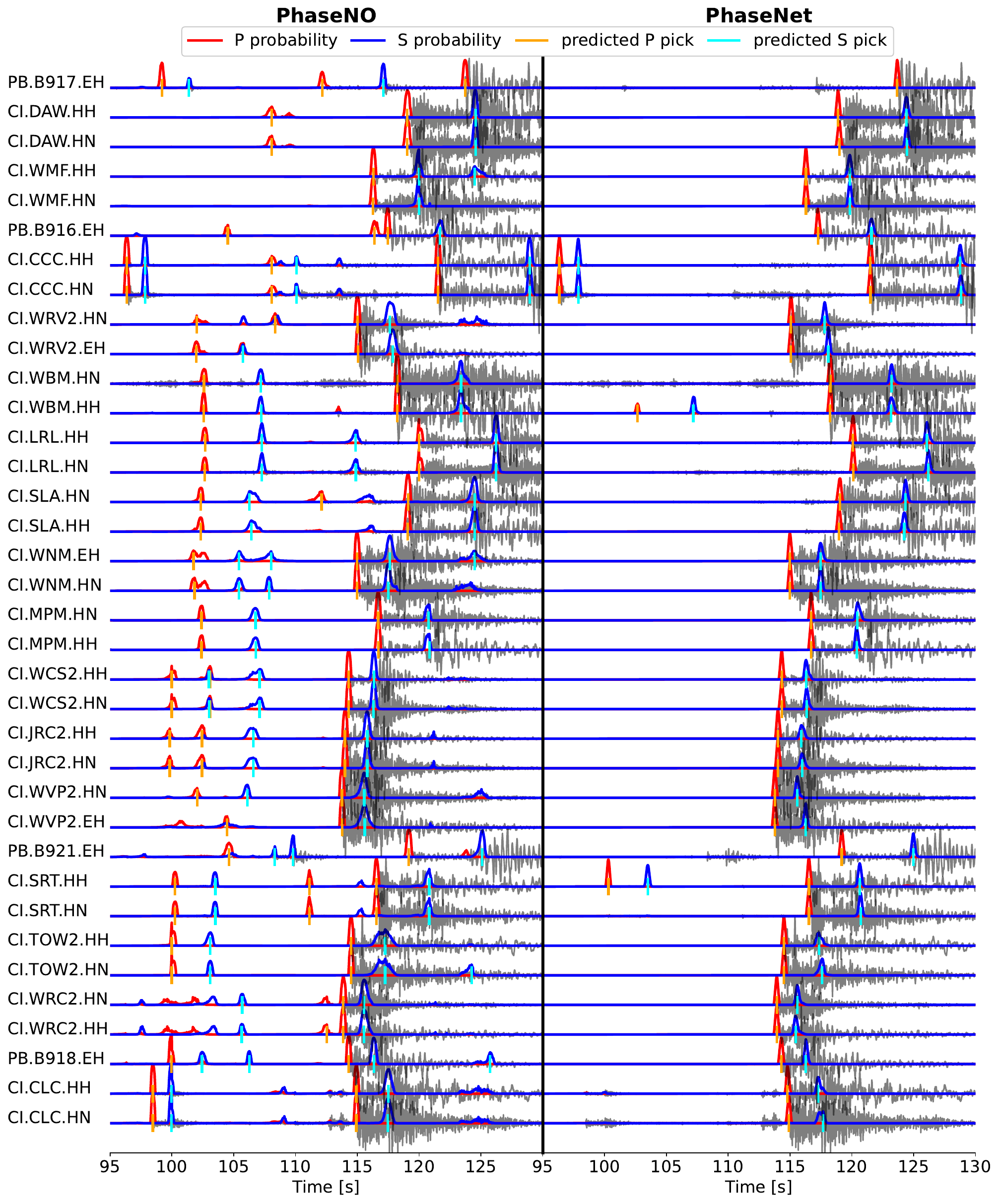}
\caption{\textbf{Representative waveforms from the 2019 Ridgecrest earthquake sequence.} We compare PhaseNO and PhaseNet on a 35-s time window (95 – 130 s) selected from one-hour continuous data starting at 06:00:00 on July 6, 2019. The picking threshold is 0.3.} 
\label{figs6:ridgecrest_waveform_95}
\end{figure}

\begin{figure}
\centering
\includegraphics[width=\textwidth]{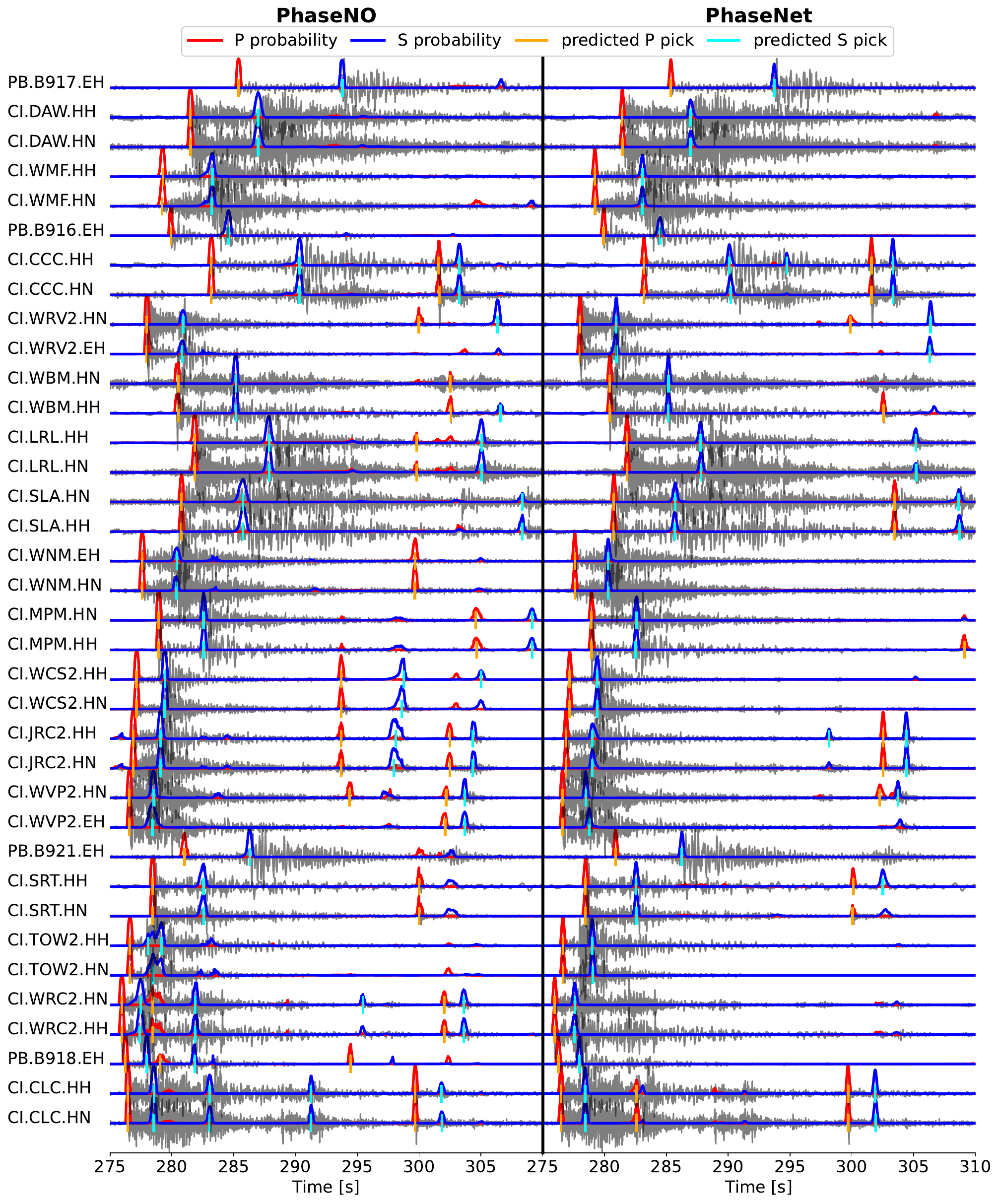}
\caption{\textbf{Representative waveforms from the 2019 Ridgecrest earthquake sequence.} We compare PhaseNO and PhaseNet on a 35-s time window (275 – 310 s) selected from one-hour continuous data starting at 06:00:00 on July 6, 2019. The picking threshold is 0.3.}
\label{figs7:ridgecrest_waveform_275}
\end{figure}

\begin{figure}
\centering
\includegraphics[width=\textwidth]{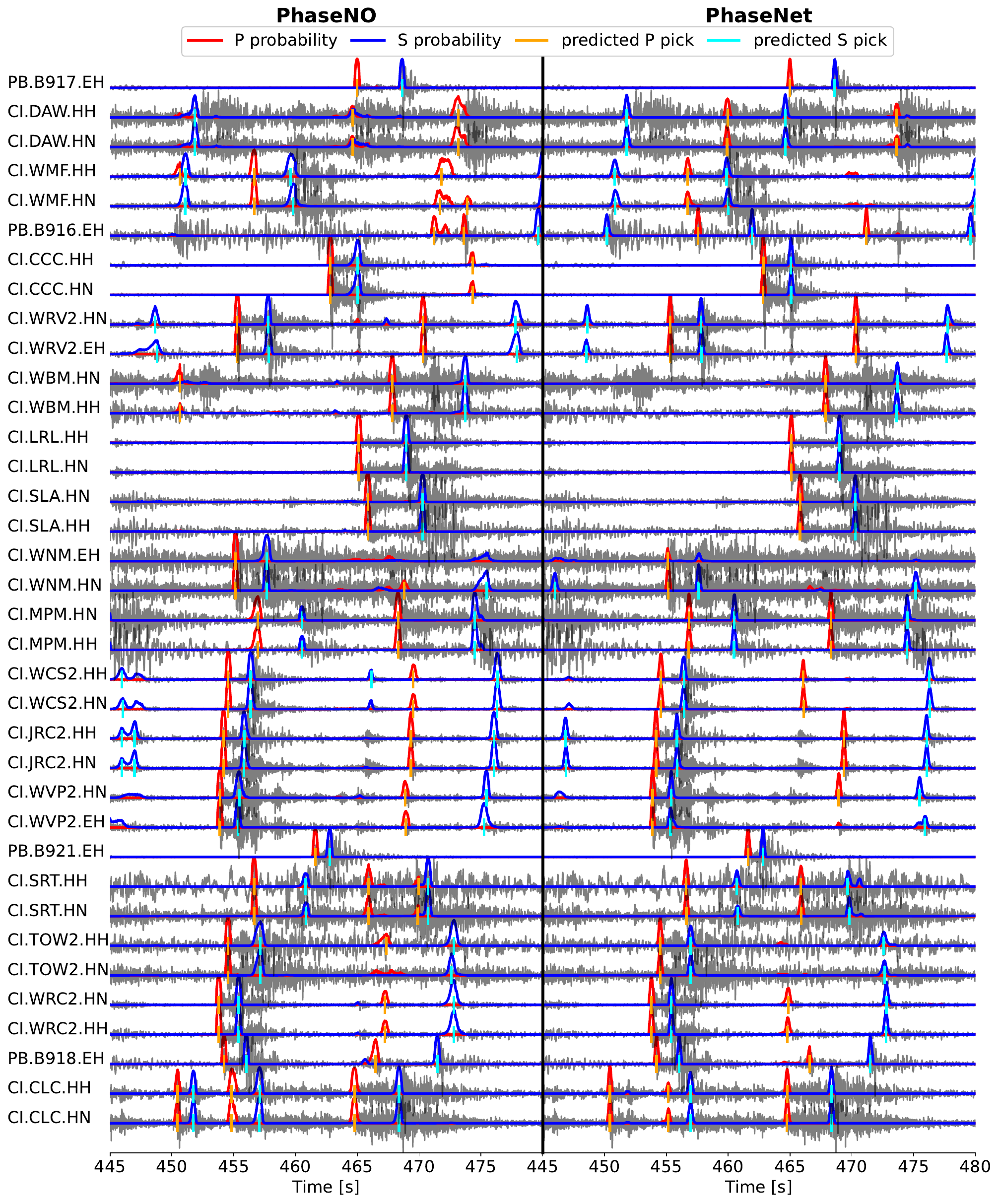}
\caption{\textbf{Representative waveforms from the 2019 Ridgecrest earthquake sequence.} We compare PhaseNO and PhaseNet on a 35-s time window (445 – 480 s) selected from one-hour continuous data starting at 06:00:00 on July 6, 2019. The picking threshold is 0.3.}
\label{figs8:ridgecrest_waveform_445}
\end{figure}

\begin{figure}
\centering
\includegraphics[width=\textwidth]{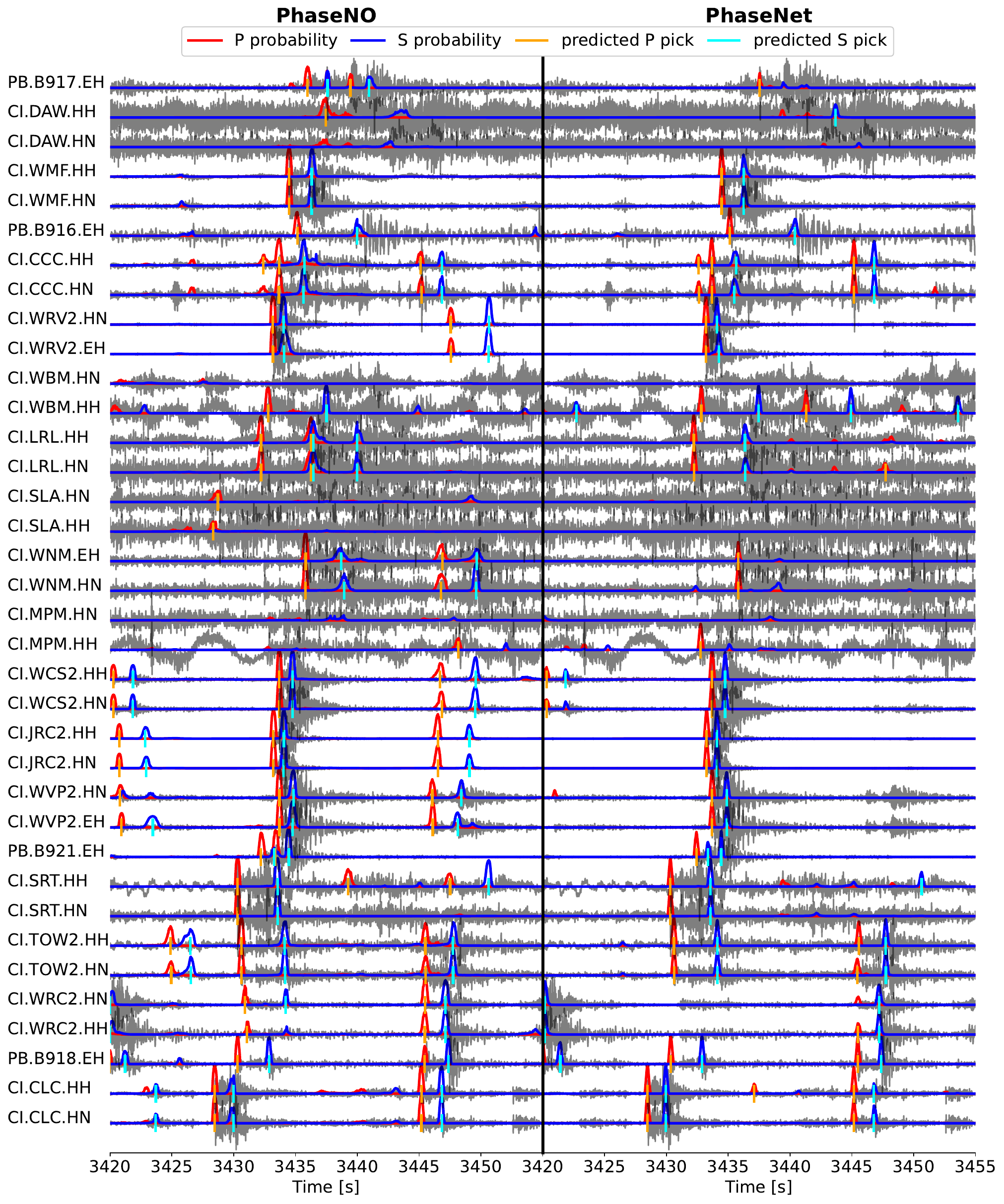}
\caption{\textbf{Representative waveforms from the 2019 Ridgecrest earthquake sequence.} We compare PhaseNO and PhaseNet on a 35-s time window (3420 – 3455 s) selected from one-hour continuous data starting at 02:00:00 on July 7, 2019. The picking threshold is 0.3.}
\label{figs9:ridgecrest_waveform_070702_3420}
\end{figure}

\begin{figure}
\centering
\includegraphics[width=0.98\textwidth]{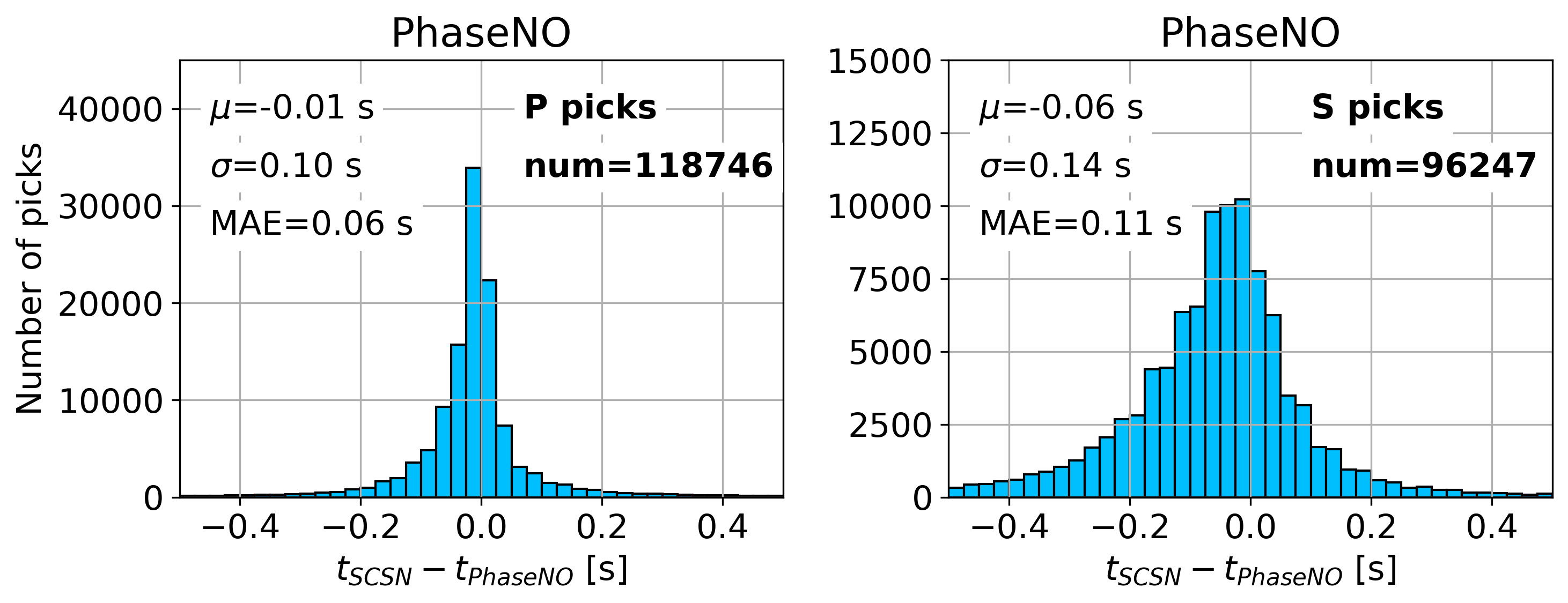}
\includegraphics[width=0.98\textwidth]{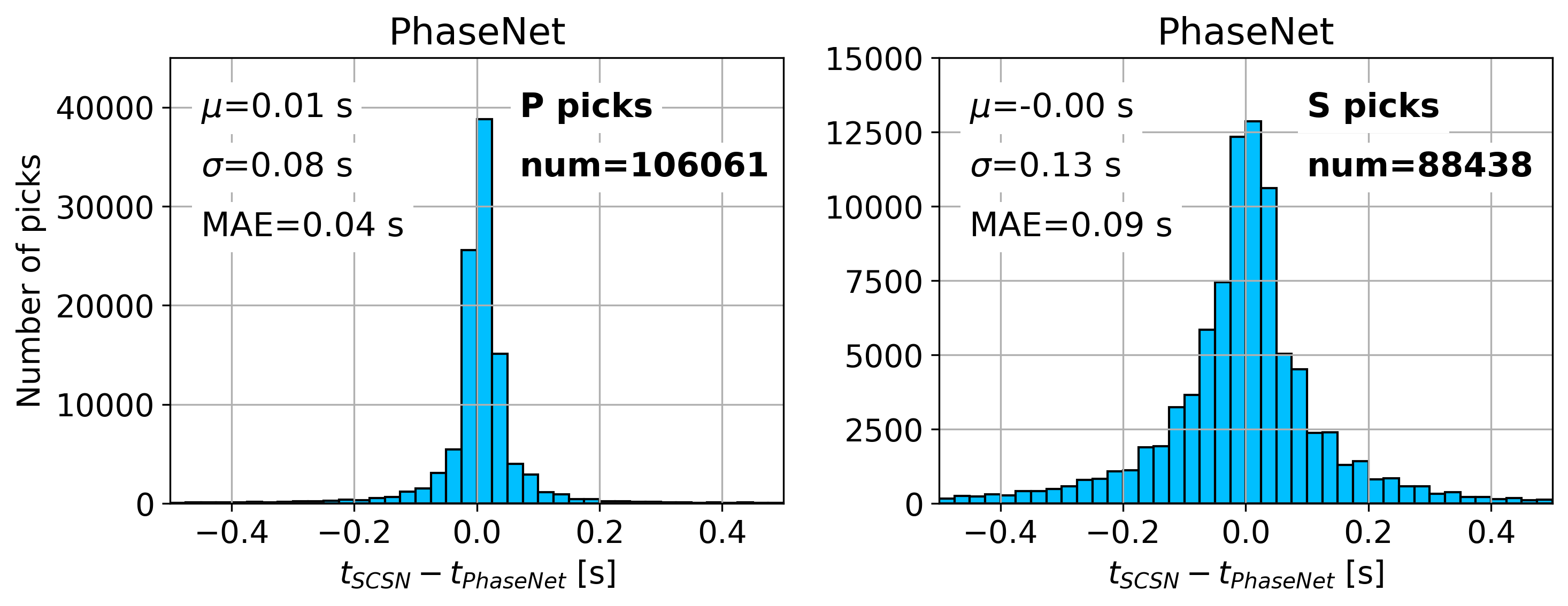}
\caption{\textbf{Travel time difference between SCSN picks and picks detected by PhaseNO or PhaseNet in the 2019 Ridgecrest earthquake sequence.} The number of picks (num) in each histogram is labeled on the panel.}
\label{figs10:ridgecrest_picks}
\end{figure}

\begin{figure}
\centering
\includegraphics[width=0.8\textwidth]{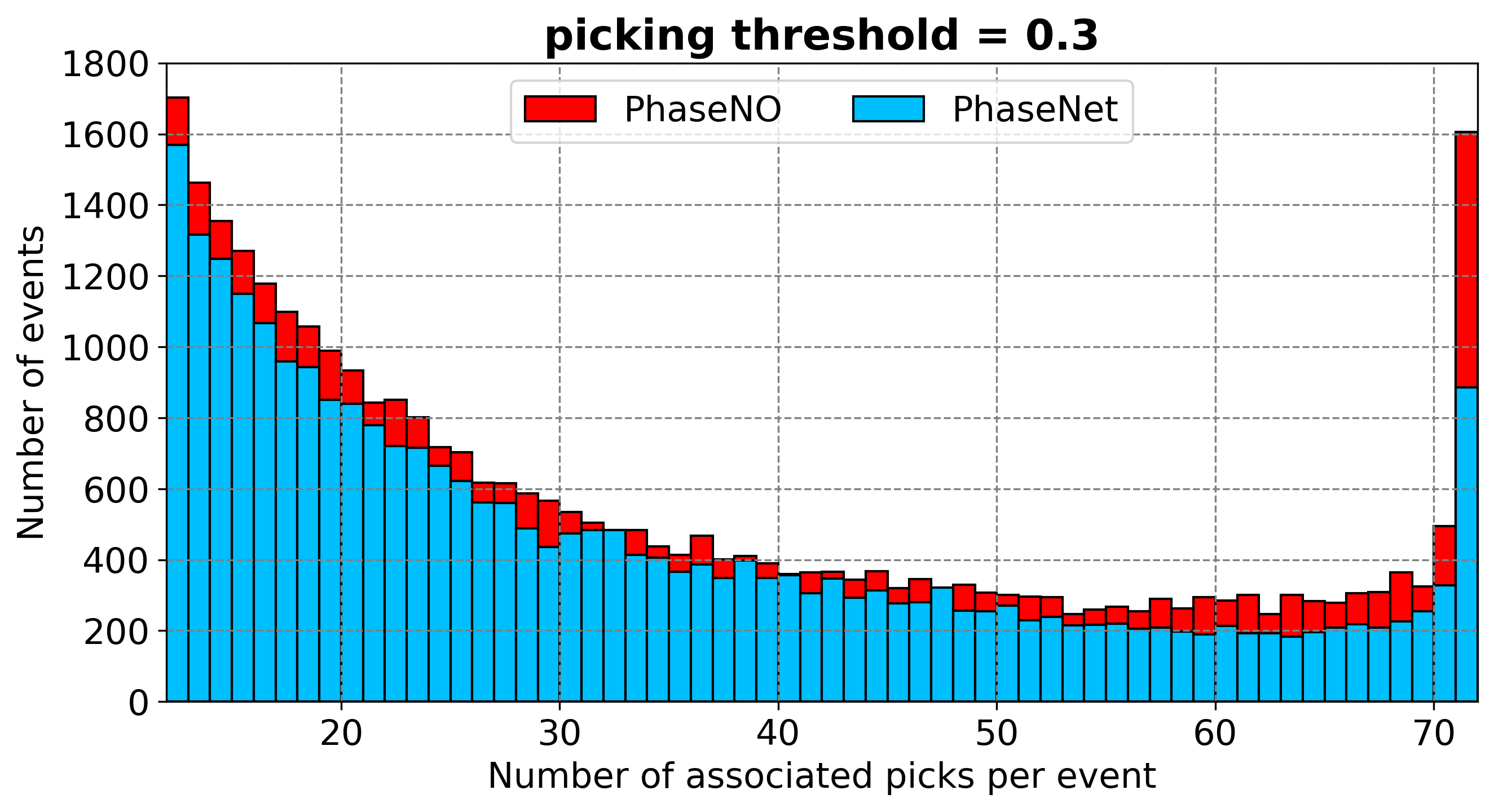}
\includegraphics[width=0.8\textwidth]{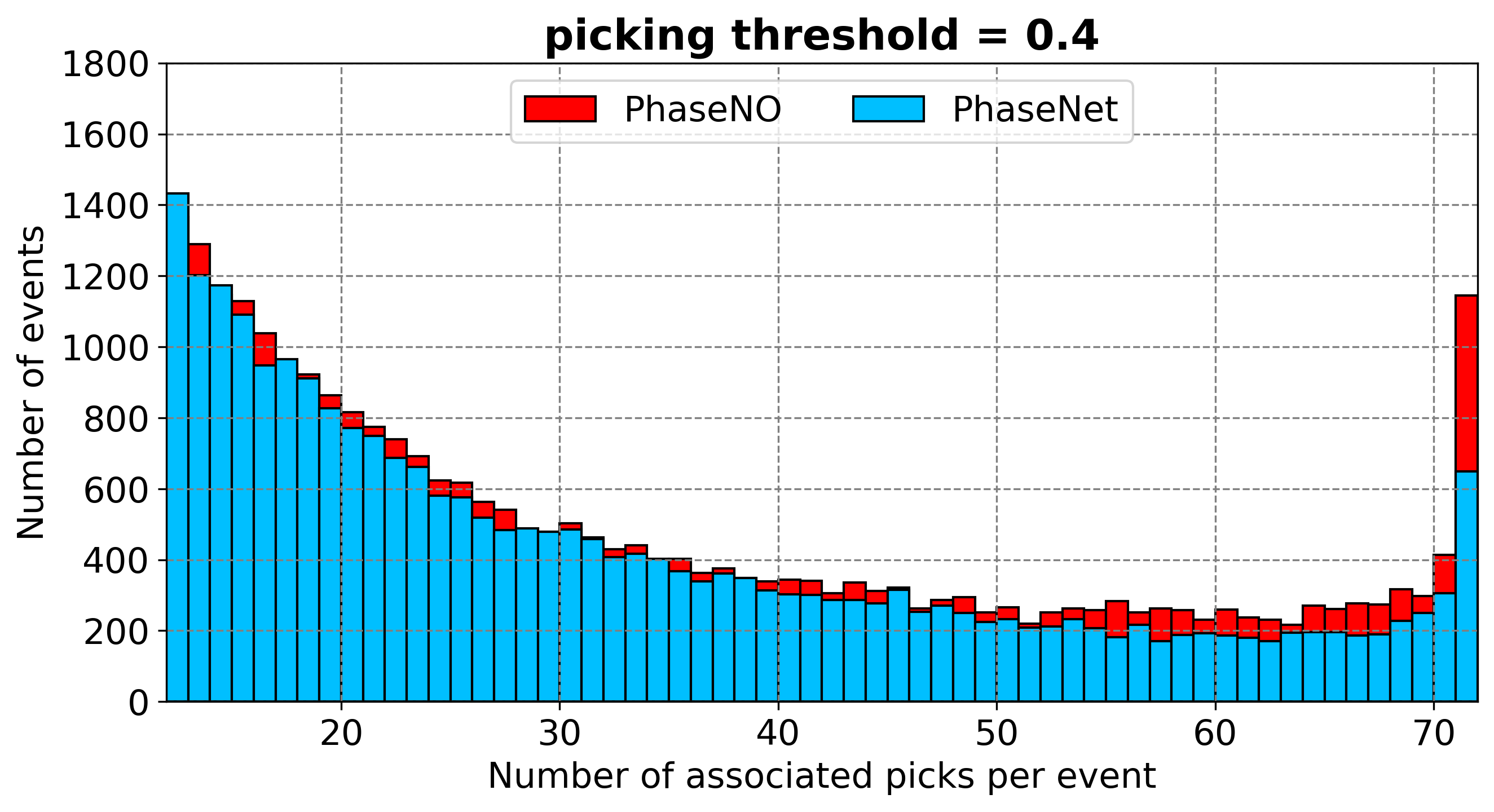}
\caption{\textbf{Comparison of the association quality of GaMMA using picks detected by PhaseNO and PhaseNet in the 2019 Ridgecrest earthquake sequence through the number of associated picks in one event.} The maximum number of the associated picks per event is 72 with 36 nodes used in this study (Table \ref{tab:stations}). The minimum is a hyperparameter pre-determined in GaMMA to filter out low quality associations. Table \ref{tab:association} compares the total number of events in a catalog with a minimum of 12, 15, and 17 in GaMMA.}
\label{figs11:ridgecrest_events}
\end{figure}

\begin{figure}
\centering
\includegraphics[width=\textwidth]{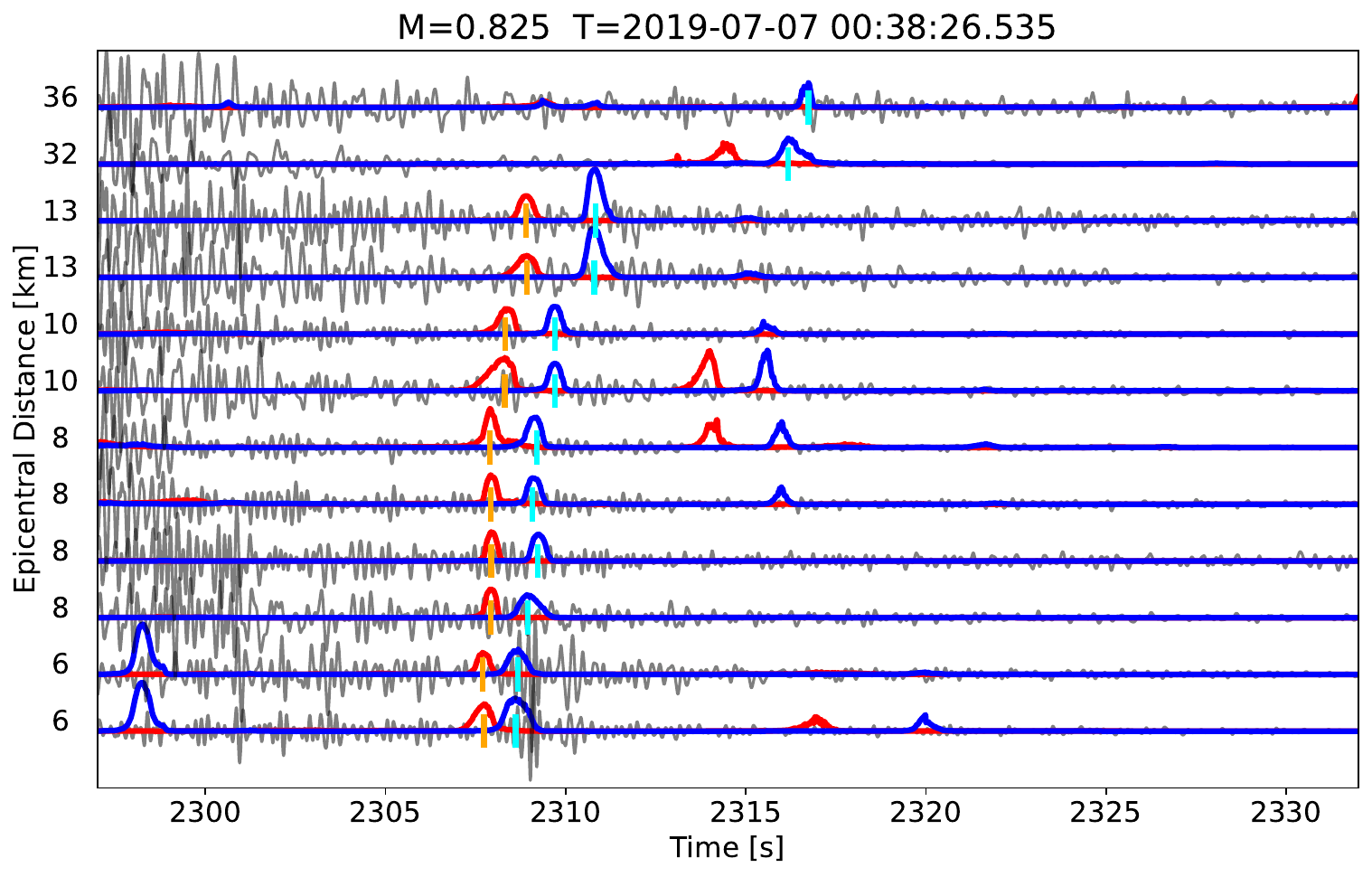}
\caption{\textbf{An example of an M 0.825 event detected by PhaseNO.} The hypocenter determined by GaMMA is 36.068, -117.843 with a depth of 2698 m. 10 P phases (orange bars) and 12 S phases (cyan bars) picked with a threshold of 0.3 were associated with this event by GaMMA. This event exists in the template matching catalog by Ross et al. but not in other catalogs.}
\label{figs12:phaseno_catalog_event_11262}
\end{figure}

\begin{figure}
\centering
\includegraphics[width=\textwidth]{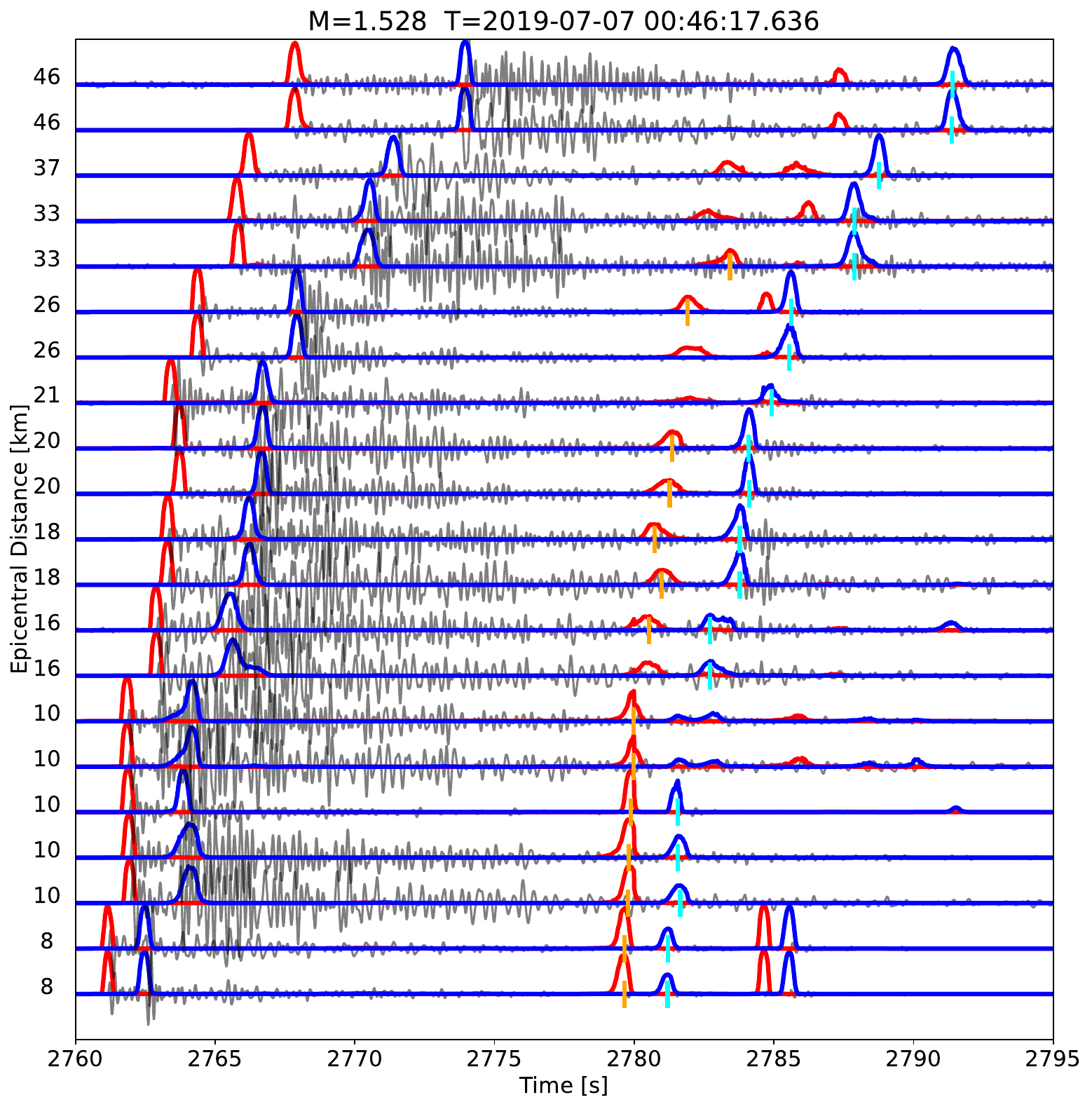}
\caption{\textbf{An example of an M 1.528 event detected by PhaseNO.} Red and blue lines are the probabilities predicted by PhaseNO on continuous data. With a threshold of 0.3, PhaseNO detects 14 P- and 19 S- phases indicated by orange and cyan vertical bars, respectively. The hypocenter determined by GaMMA is 35.862, -117.671 with a depth of 8059 m. This event exists in the template matching catalog by Ross et al. but not in other catalogs. This event is difficult to detect using PhaseNet, probably because the first arrivals overlap with the coda of the larger event.}
\label{figs13:phaseno_catalog_event_11293}
\end{figure}

\begin{figure}
\centering
\includegraphics[width=\textwidth]{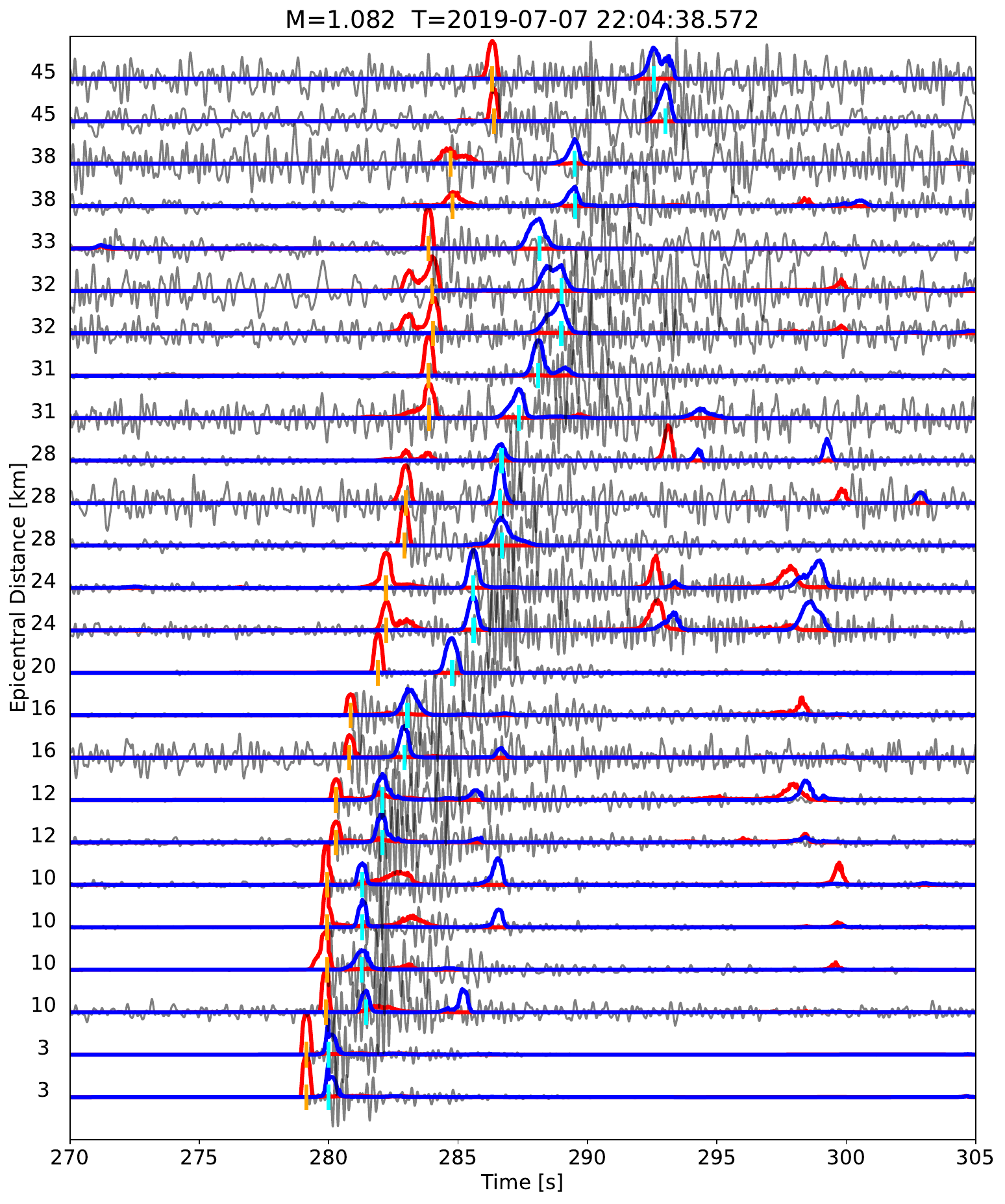}
\caption{\textbf{An example of an M 1.082 event detected by PhaseNO.} The hypocenter determined by GaMMA is 36.089, -117.849 with a depth of 295 m. 24 P phases (orange bars) and 25 S phases (cyan bars) picked with a threshold of 0.3 were associated to this event by GaMMA. This event was detected by both machine-learning pickers (PhaseNO and PhaseNet) but did not exist in neither the template matching catalogs or the SCSN catalog.}
\label{figs14:phaseno_catalog_event_16248}
\end{figure}

\begin{figure}
\centering
\includegraphics[width=0.8\textwidth]{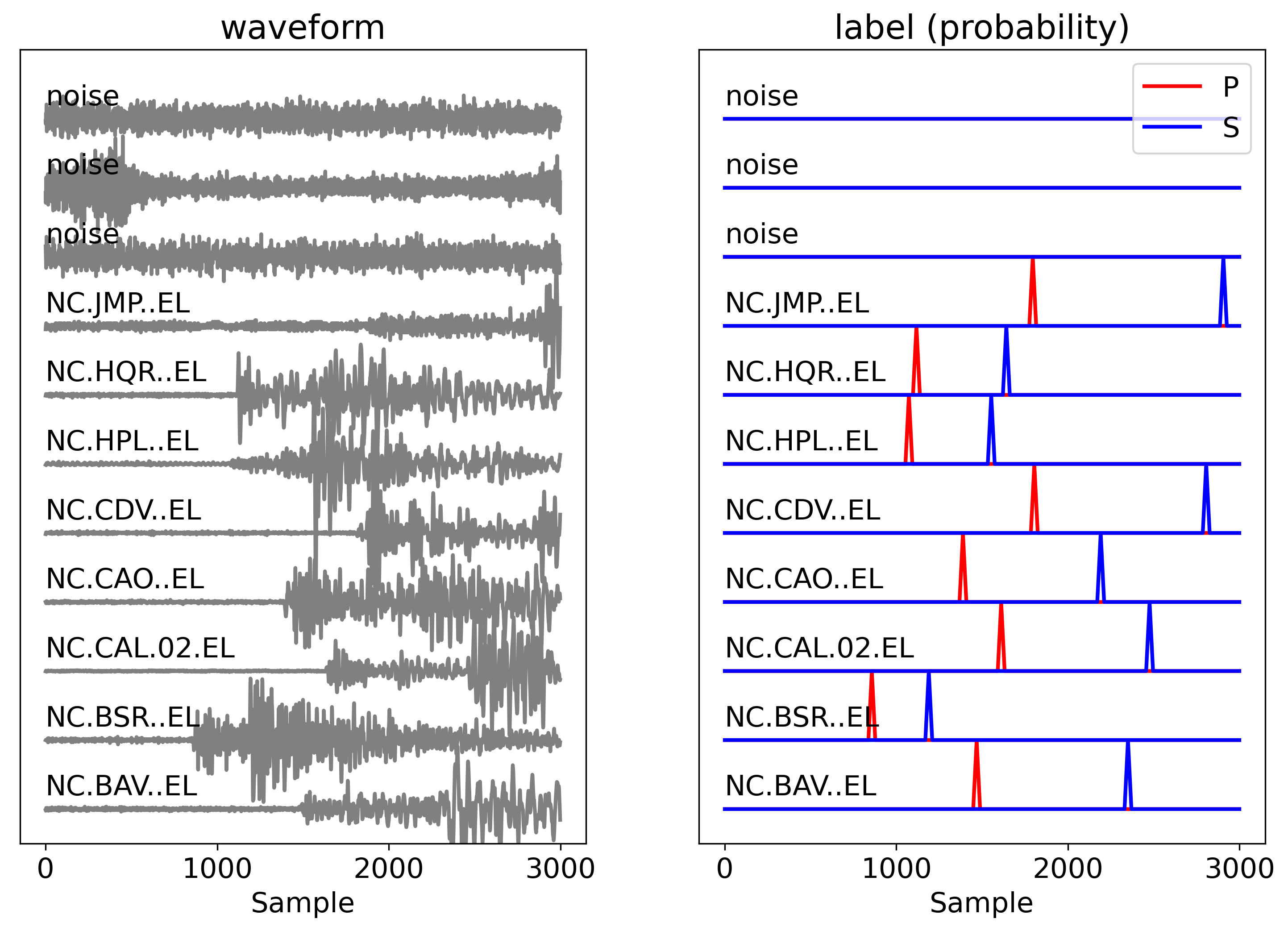}
\includegraphics[width=0.6\textwidth]{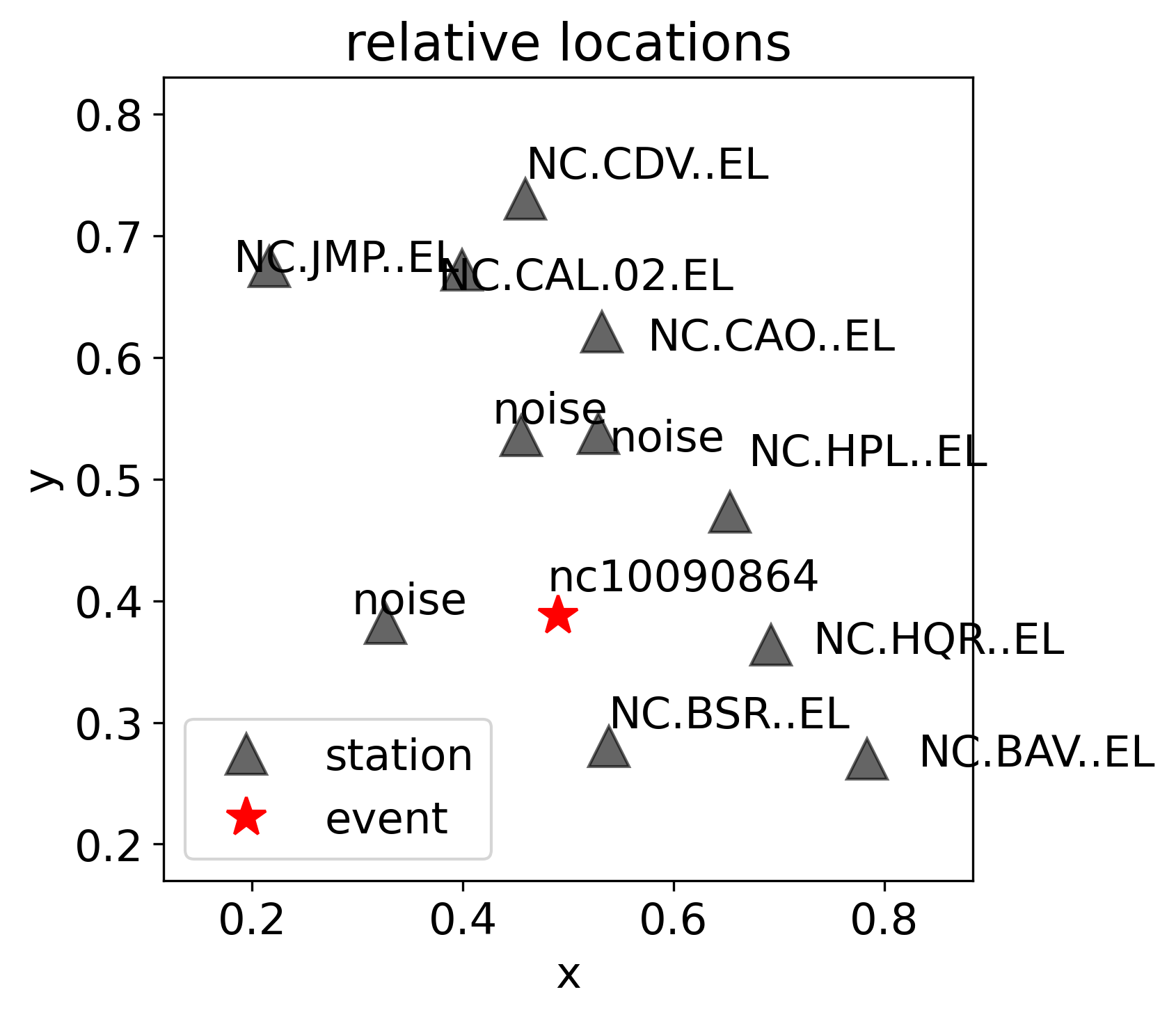}
\caption{\textbf{Example of one graph-type sample containing one event from the training dataset.} The sample records one event with eight stations. Additionally, three virtual stations with only noise waveforms are randomly placed in the graph. In total, the graph contains 11 nodes.}
\label{figs17:graph_example1}
\end{figure}

\begin{figure}
\centering
\includegraphics[width=0.8\textwidth]{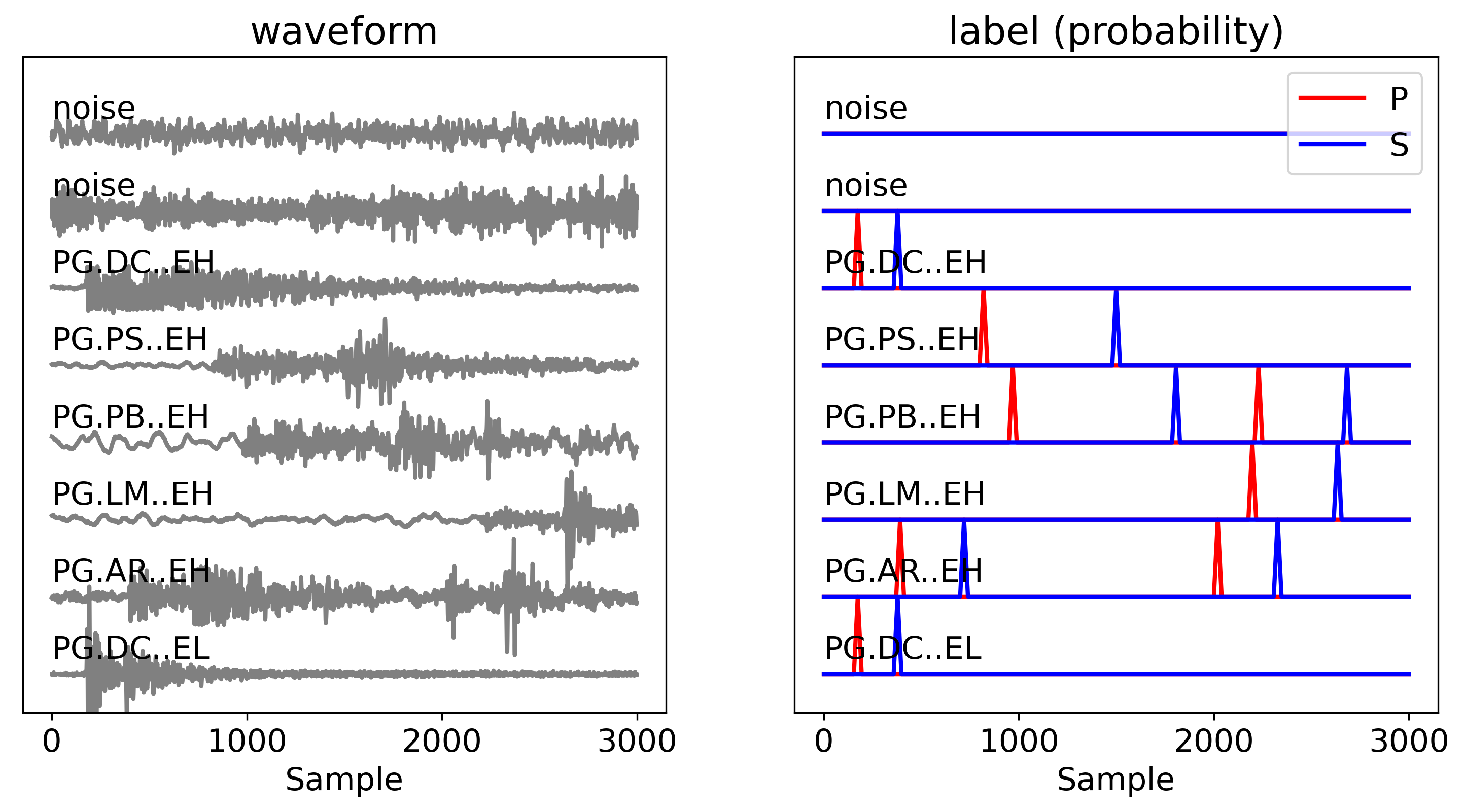}
\includegraphics[width=0.5\textwidth]{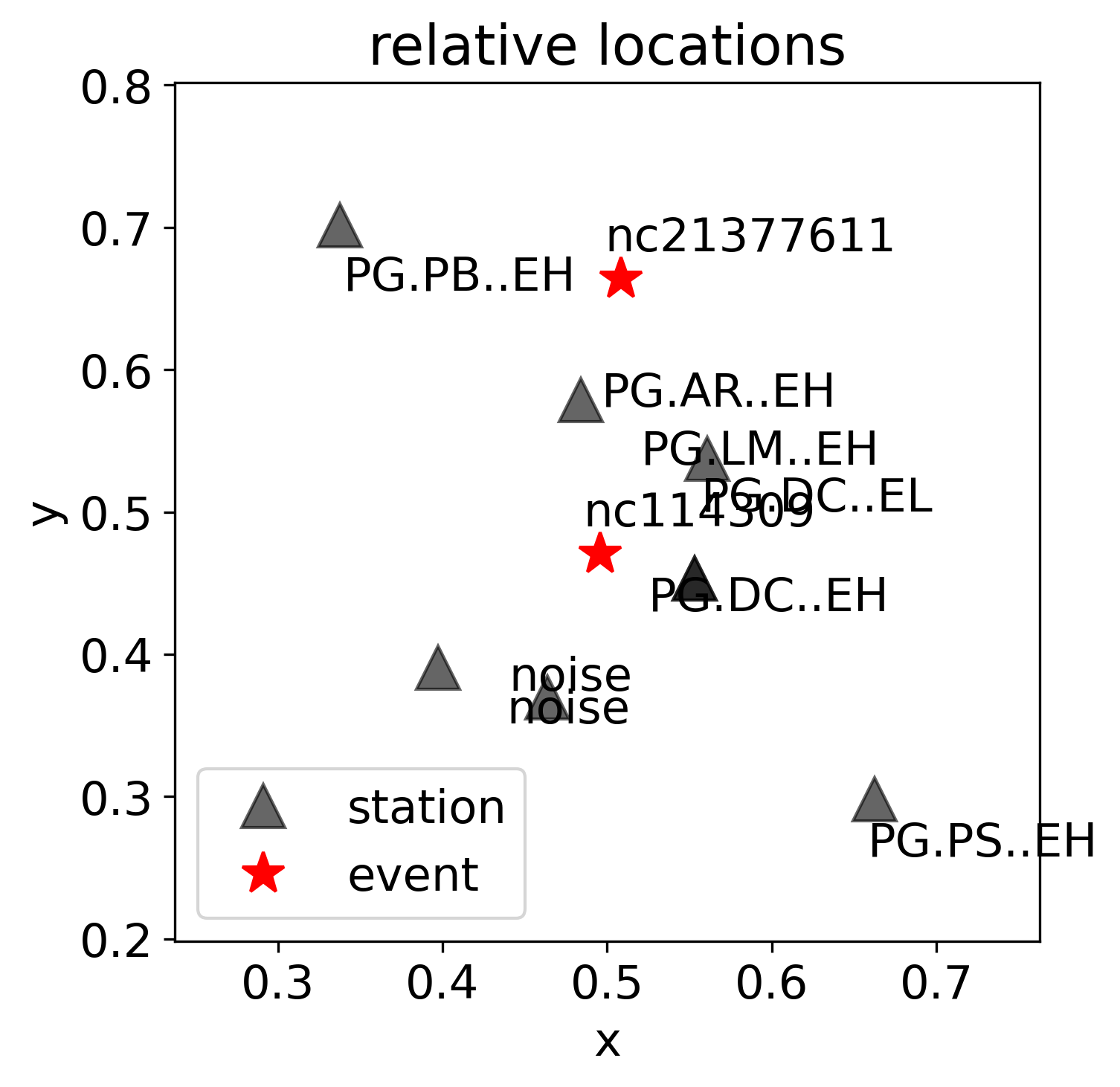}
\caption{\textbf{Example of one graph-type sample containing two events from the training dataset.} The sample records two events by six nodes. Nodes PG.PB..EH and PG.AR..EH record the waveforms of both events. Other nodes contain a single event. Two virtual nodes with noise waveforms and random locations are added to the graph. Stations PG.DC..EH and PG.DC..EL are considered as two nodes at the same location. In total, the graph contains 8 nodes.}
\label{figs18:graph_example2}
\end{figure}

\begin{figure}
\centering
\includegraphics[width=0.8\textwidth]{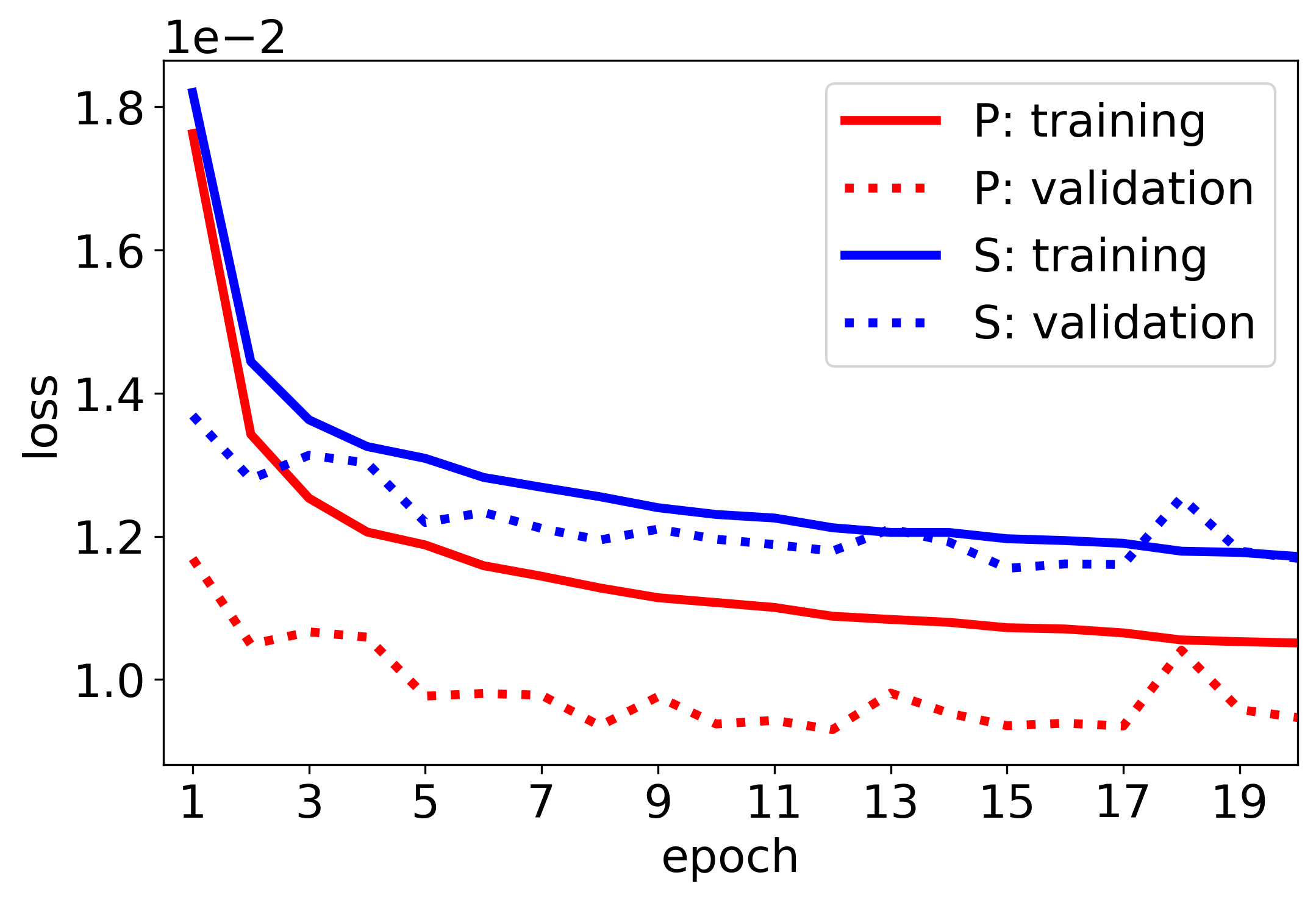}
\caption{\textbf{Learning curves of PhaseNO.} The loss of S phase is higher than that of P phase, implying that S phase picking is more challenging than P phase.}
\label{figs19:loss}
\end{figure}

\begin{figure}
\centering
\includegraphics[width=\textwidth]{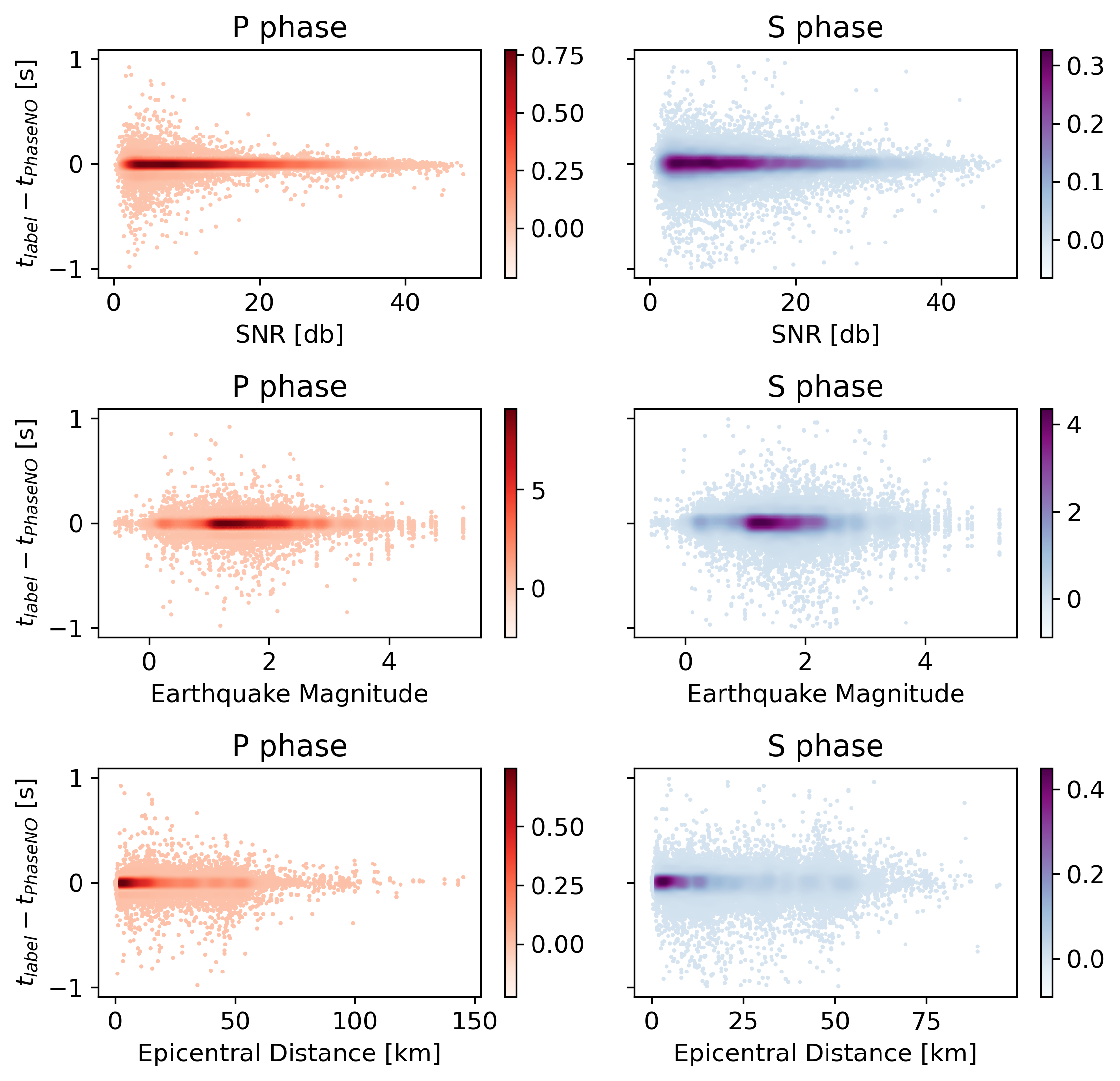}
\caption{\textbf{Prediction errors as a function of SNR, earthquake magnitude, and epicentral distance.} The colormap shows the point density. Prediction errors are computed as the travel time difference between predicted phases and manually picked phases (label). Only when the prediction error is less than 0.5 s did we consider the prediction as true positive.}
\label{figs15:factors_errors}
\end{figure}

\begin{figure}
\centering
\includegraphics[width=\textwidth]{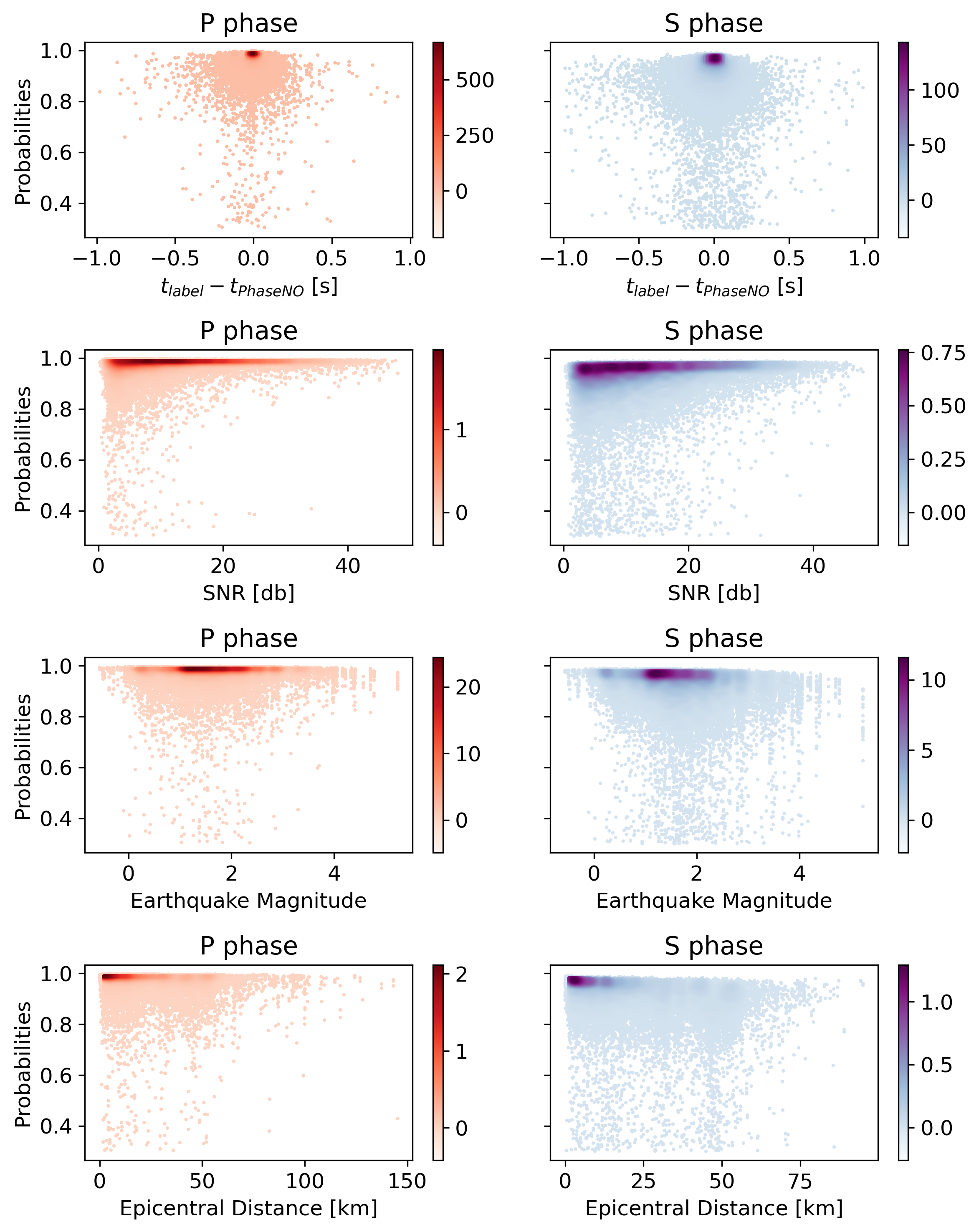}
\caption{\textbf{Predicted probabilities as a function of prediction errors, SNR, earthquake magnitude, and epicentral distance.} The colormap shows the point density.}
\label{figs16:factors_probs}
\end{figure}

\end{document}